\RequirePackage{fix-cm}
\documentclass{article}
\usepackage[numbers]{natbib}
\usepackage{setspace}
\usepackage[fleqn]{amsmath}
\usepackage{multirow}
\usepackage{graphicx}
\usepackage{subfigure}
\usepackage{stmaryrd}
\usepackage{xcolor}
\usepackage{amssymb, amstext}
\usepackage{bm}     % bold math symbols via \bm
\usepackage{bbm}    % characteristic function via \mathbbm{1}

\providecommand{\keywords}[1]{\noindent \textbf{Keywords:} #1}

\begin{document}
\title{Stochastic aspects of crack deflection and crack path prediction in short fiber reinforced polymer matrix composites}
%\titlerunning{Stochastic aspects of crack paths}        % if too long for running head
\author{A. Ricoeur$^{a,\dagger}$, F. Lindner$^{b}$, K. Zarjov$^{a}$\\
{\small{$^{a}$Institute of Mechanics, University of Kassel, 34125 Kassel, Germany}} \\
{\small{$^{b}$Institute of Mathematics, University of Kassel, 34125 Kassel, Germany}} \\
{\small{$^{\dagger}$Corresponding author: ricoeur@uni-kassel.de}}}
\date{}
\maketitle
\begin{abstract}
Owing to the production process, short fiber reinforced composites exhibit a pronounced anisotropy of both elastic properties and crack growth resistance.
In particular the latter issue has a major impact on crack deflection and inevitably has to be taken into account for the sake of an accurate prediction of crack paths.
The perpendicular axes of transverse isotropy are associated with the fiber orientations, whereupon a crack in transverse direction encounters the largest fracture toughness.
While the local mean fiber orientations in polymer matrix composites are determined by the injection molding process, their statistical fluctuations can approximately be described in terms of Gaussian random field models.
Furthermore, statistical variations of the volume fraction of fibers and the fiber--matrix adhesion give rise to stochasticity of the local ratio of fracture toughness anisotropy.
Influencing factors of prediction regions of crack paths obtained from finite element simulations with an adapted J-integral deflection criterion are investigated, just as stochastic aspects of bifurcation phenomena of crack deflection observed under mode-I loading.
\end{abstract}


\newpage

\section{Introduction}
\label{sec:intro}
Fiber reinforced composites provide mechanical properties in terms of strength, reliability and life span comparable to conventional engineering materials, however, with much less specific weight.
The application of short fibers, in particular, in conjunction with a polymer matrix enables the production of a variety of structural components of vessels, automobiles, space- and aircrafts by injection molding, being an efficient, flexible and economic manufacturing technique.

Focusing on strength and lifetime, the understanding of damage and fracture behaviors plays a crucial role in the reliable assessment of structural components.
In contrast to classical, macro- and mesoscopically homogeneous engineering materials, the associated processes are more complex, involving, inter alia, cracking of the matrix, delaminations of fibers and matrix, or the rupture of fibers.
With regard to the latter aspect, cellulosic fibers are favorable, e.g. compared to glass fibers, due to their comparably high strength and elongation at rupture, giving rise to distinctly improved impact strength and less spontaneous failure of the composite \cite{feld1,feld2}.

While microcracking is responsible for a local initiation of damage, lifetime and reliability of the structural component finally are determined by the growth of macroscopic cracks.
The lengths of these cracks and the fibers typically differ by at least one order of magnitude, thus a scale separation is feasible, considering the problem of a crack in a homogenized effective material.
The constitutive behavior of the material is determined by the microstructural features outlined above and is measured with specimens, typically containing millions of fibers \cite{judt2}.

In this context, the fracture toughness plays a crucial role, essentially depending on the orientations of fibers at the crack tip.
The injection molding process evokes a predominant direction (PD), roughly aligning fibers parallel to the flow front, which is approximately parabolic.
Consequently, anisotropy of the fracture toughness and crack growth resistance, respectively, is observed, whereupon crack growth perpendicular to the PD is impeded.
In the case of regenerated cellulose fiber (RCF) reinforced polypropylene (PP), the fracture toughness in PD and transverse direction (TD) differs by a factor 1.45-1.7, depending on weight ratio and coupling agent \cite{judt2}.

In \cite{judt2, judt1}, crack paths in RCF and glass fiber (GF) reinforced PP have been investigated both experimentally and theoretically on the basis of compact tension (CT) -specimens.
It was found that the prerequisites of linear elastic fracture mechanics in terms of small scale yielding are satisfied and that an extended $J$-integral criterion is suitable for predicting crack deflection numerically.
Although the alignment of fibers effectuates a transverse isotropy of elastic coefficients just as of the crack growth resistance, the influence of directional elasticity on crack deflection turned out to be negligible.
Analytical calculations further yielded an interesting bifurcation feature of crack growth under mode-I loading, whereupon a crack may deflect in either direction if fibers at its tip are perpendicular to the crack and a critical ratio of anisotropy is exceeded.

While in \cite{judt2, judt1} the orientations of fibers and all related parameters are assumed to be deterministic, i.e. fibers being uniformly aligned along a PD and ratios of anisotropy being constant in the whole specimen, stochastic aspects of a real composite are considered at this point.
Crack paths taken from experiments with different specimens of identical geometry indicate the non-deterministic nature of the problem, basically exhibiting a scatter for reproduced loading. Conditions leading to the above mentioned bifurcation may occur randomly at any position during crack growth where mixed-mode loading prevails, in some cases leading to a pronounced deflection from the scatter range.
Numerical simulations with the finite element method (FEM) and semi-analytical considerations of deflecting cracks, based on stochastic modeling of fiber-related and location-dependent macroscopic quantities, foster the understanding of experimental crack paths and improve the quality of prediction.
The employed stochastic model parameters are partly motivated by micro computed tomography ($\mu$--CT) of RCF reinforced PP.

Crack growth simulations in anisotropic solids based on classical fracture mechanics discontinuity assumptions, i.e. introducing internal free surfaces as crack faces meeting at a sharp crack tip, are e.g. reported in \cite{boone,chen,gao17,hakim,judt4}.
The directions of incremental crack advances are predicted based on a deflection criterion.
Commonly applied criteria are the maximum tangential stress (MTS) criterion \cite{erdo}, the maximum strain energy density (SED) criterion \cite{sih74}, the maximum energy release rate criterion \cite{hussa,nuis} and the $J$-vector criterion \cite{ma05,stri}.
For orthotropic materials, the MTS criterion has been extended by Saouma et al. \cite{saou}, where modifications are required to account for both elastic and fracture mechanics anisotropies.
The $J$-vector criterion intrinsically holds for arbitrary elastic anisotropy and an extension towards direction-dependent fracture toughness has been introduced and verified experimentally by Judt et al. \cite{judt4}.
Carloni and Nobile \cite{carlo} generalized the SED criterion for orthotropic solids.
Interpolation functions applied to experimental findings of typically two or three crack directions in a specimen provide continuous analytic expressions for the angular dependence of the fracture toughness \cite{carl03,judt4,kfo96,saou}.
Stochastic aspects of fracture mechanics refer to fatigue crack growth rates, e.g. in \cite{riahi10,salimi18,yang96}, whereas crack deflection and thus paths remain deterministic.
Gerasimov et al.\ \cite{gerasimov} recently investigated stochastic crack paths resulting from random perturbations of energy functionals within the phase field approach to brittle fracture, focusing on a primarily isotropic two-dimensional setup with a single circular hole.

To our best knowledge, numerical fracture-mechanical simulations or (semi-)analytical investigations of crack paths in short fiber reinforced composites or other anisotropic solids do not yet take into account stochastic aspects associated with fiber orientations or distributions and crack deflection.

\section{Theoretical fundamentals}
\label{sec:theo}
\subsection{Crack tip loading and deflection} \label{ssec:crack}

\begin{figure}[!ht]
\begin{centering}
\includegraphics[width=.5\textwidth]{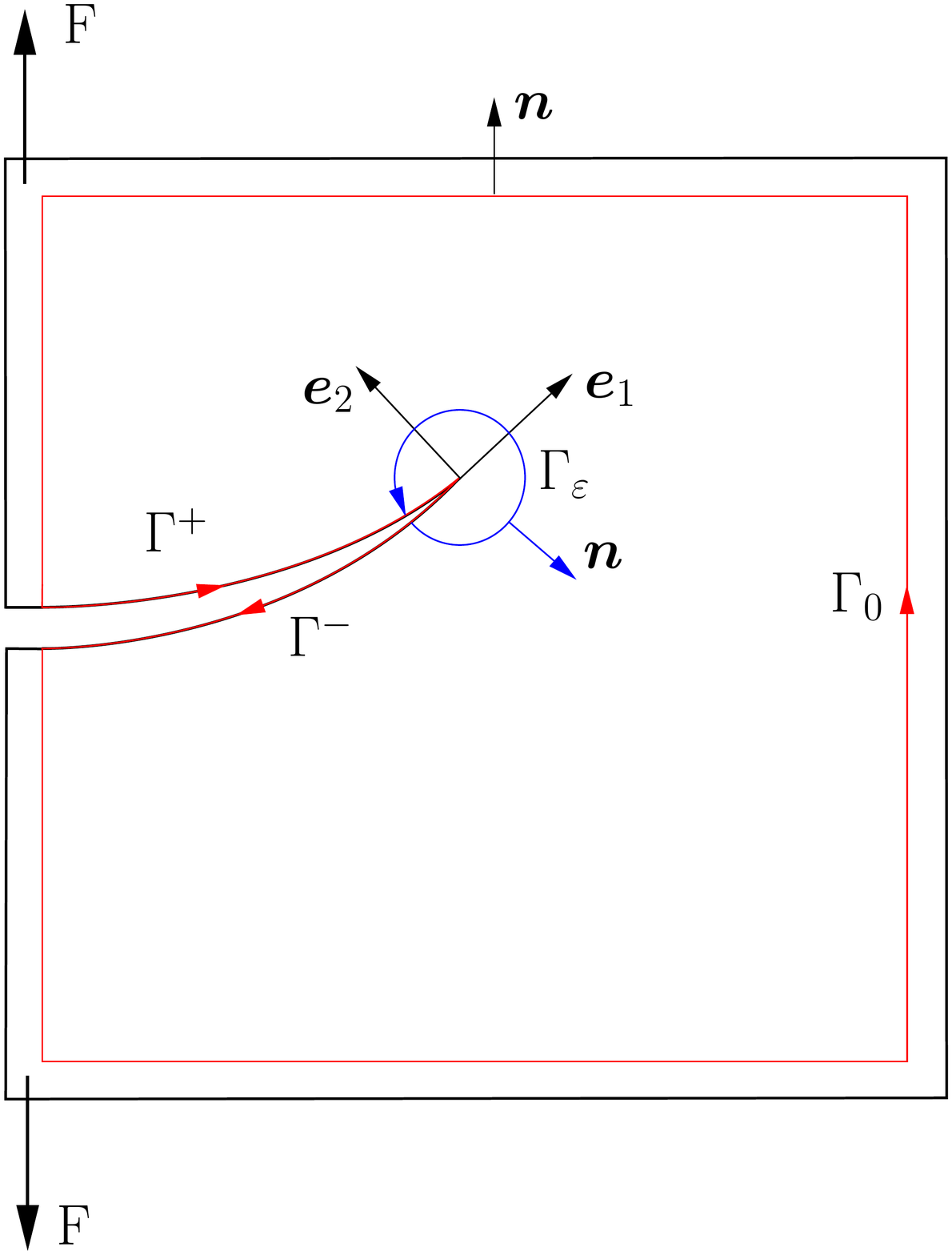}
\par\end{centering}
\protect\caption{Curved crack faces and integration paths $\Gamma_0+\Gamma^{+}+\Gamma^{-}$ and $\Gamma_{\varepsilon}$ of the $J$-integral with unit normal $\mathbf{n}$ and crack tip coordinate system $(\mathbf{e}_1,\mathbf{e}_2)$}
\label{Fig11}
\end{figure}

A curved crack, exposed to in-plane mixed-mode loading at its tip, is considered according to Fig.~\ref{Fig11}, where a quadratic plate, loaded by a pair of forces, reminds of a CT-type specimen.
The mixed-mode ratio is defined as
\begin{equation} \label{eq:1}
\Phi=\arctan\left(\frac{K_{II}}{K_{I}}\right)\ ,
\end{equation}
where $K_I$ and $K_{II}$ as usual denote the stress intensity factors.
The $J$-integral is employed for crack tip loading analyses and with an infinitely small contour $\Gamma_{\varepsilon}$ with outward unit normal $n_j$ it reads \cite{budric, cherep}
\begin{equation} \label{eq:2}
J_k=\lim_{\varepsilon \to 0}\int \limits_{\Gamma_\varepsilon} Q_{kj} n_j \mathrm{d}S=\lim_{\varepsilon \to 0}\int
\limits_{\Gamma_\varepsilon}\left(u\delta_{kj}-\sigma_{ij}u_{i,k}\right)n_j\mathrm{d}S\ .
\end{equation}
The analytical notation is applied with lower case indices taking values 1 to 3 in a general three-dimensional problem, and repeated indices imply summation.
Further, a comma denotes a spacial derivative, e.g. in the above equation $u_{i,k}$ is the displacement gradient.
Two-dimensional crack problems are addressed in Eq.~(\ref{eq:2}), thus indices are confined to values 1 and 2.
The Eshelby tensor $Q_{kj}$ has been introduced, incorporating the stress tensor $\sigma_{ij}$ and the identity tensor $\delta_{ij}$ as well as the elastic energy density
\begin{equation} \label{eq:3}
u(\varepsilon_{ij})=\frac{1}{2}E_{ijkl}\varepsilon_{ij}\varepsilon_{kl}\ ,
\end{equation}
representing the specific potential in case of a linear elastic problem.
The stress tensor as associated variable is accordingly obtained by differentiation:
\begin{equation} \label{eq:4}
\sigma_{ij}=\frac{\partial u}{\partial \varepsilon_{ij}}=E_{ijkl}\varepsilon_{kl} = \frac{\partial^2 u}{\partial \varepsilon_{ij}\partial \varepsilon_{kl}}\varepsilon_{kl}\ .
\end{equation}
The fourth order elastic tensor with symmetry properties $E_{ijkl}=$ $E_{jikl}=$ $E_{ijlk}=$ $E_{klij}$ introduces Hooke's Law for the most general case, whereupon five independent elastic coefficients are basically required for the transversally isotropic model of fiber reinforced composites.

\begin{figure}[!ht]
\begin{centering}
\includegraphics[width=.85\textwidth]{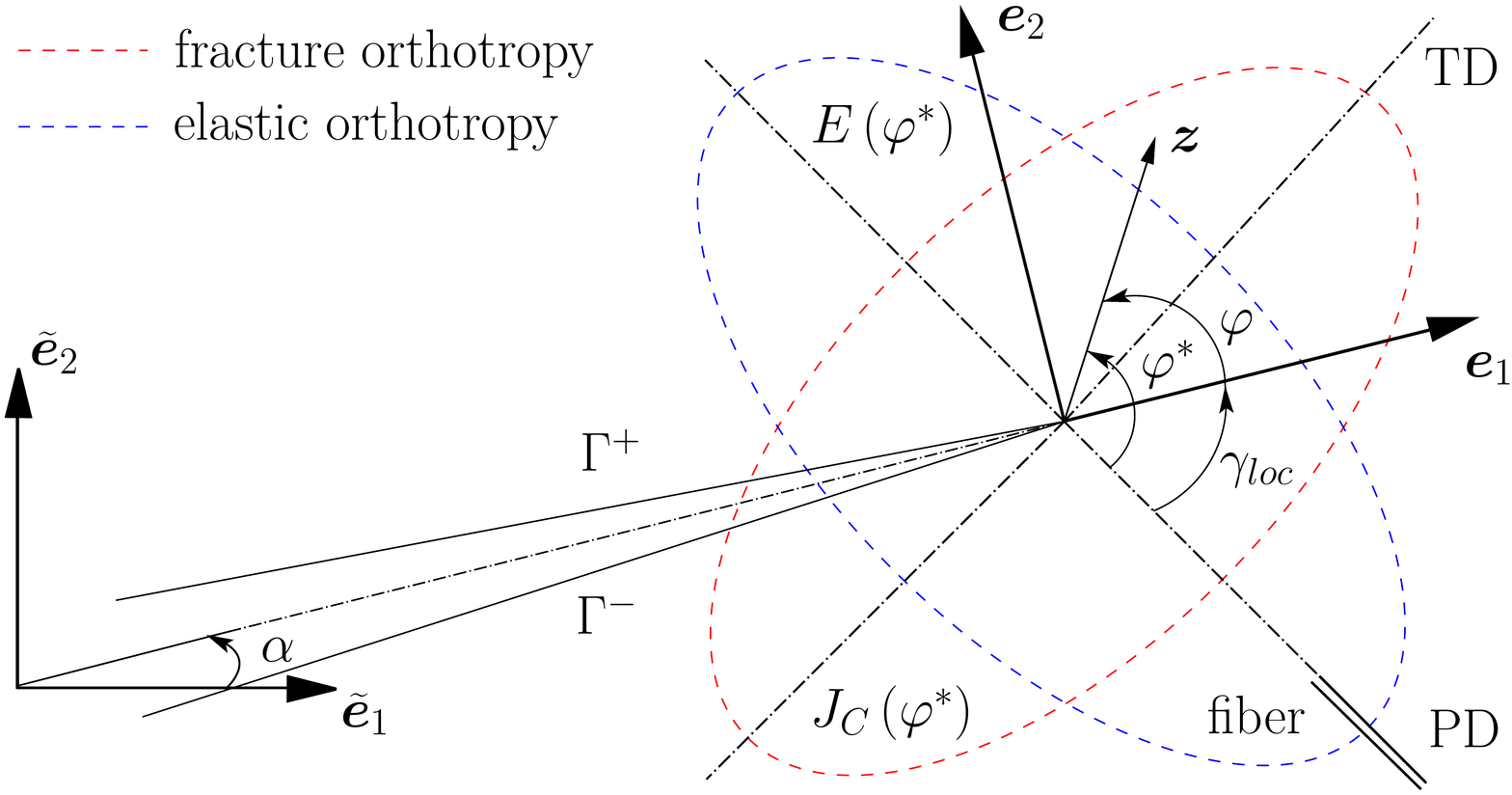}
\par\end{centering}
\protect\caption{Local crack tip and global coordinate systems $(\mathbf{e}_1, \mathbf{e}_2)$ and $(\tilde{\mathbf{e}}_1, \tilde{\mathbf{e}}_2)$; interpolation of anisotropies of elastic constants $E(\varphi^{\ast})$ and crack growth resistance $J_C(\varphi^{\ast})$ related to the fiber orientation along the predominant direction (PD); unit vector of crack deflection $\mathbf{z}$; global fiber orientation $\gamma=\gamma_{\text{loc}}-\alpha$}
\label{Fig0}
\end{figure}

In the numerical simulations of crack growth, a remote countour $\Gamma_{0}$ is taken for the $J$-integral, being complemented by contours $\Gamma^{+}$, $\Gamma^{-}$ along the positive and negative crack faces for the sake of path-independence, see Fig.~\ref{Fig11}:
\begin{equation} \label{eq:5}
J_k = \int\limits_{\Gamma_0} Q_{kj} n_j\mathrm{d}S
+ \int\limits_{\Gamma^{+}} \left\llbracket Q_{kj} \right\rrbracket^+_- n_j\mathrm{d}S\ .
\end{equation}
The squared bracket represents the jump of the energy-momentum-tensor across the crack faces, where $\Gamma^{+}=-\Gamma^{-}$ holds due to geometrical linearity.
Tractions are further assumed to vanish, thus the jump term is simplified according to
\begin{equation} \label{eq:6}
\left\llbracket Q_{kj} \right\rrbracket^+_- n_j = \left\llbracket u \right\rrbracket^+_- n_k = \frac{1}{2}\left(\sigma_{tt}^+ \varepsilon_{tt}^+ - \sigma_{tt}^- \varepsilon_{tt}^-\right)n_k \ ,
\end{equation}
where the $\sigma_{tt}$ and $\varepsilon_{tt}$ denote tangential normal stress and strain, respectively, along the crack faces.
Approaching the crack tip, the values obtained from the finite element calculations are inaccurate, partly leading to considerable errors in $J_2$, unless measures are taken \cite{eischen,judt3} to improve the accuracy of the crack surface integral.
The energy release rate of an infinitesimal crack extension into a direction indicated by the unit vector $z_k$ in a specimen with a thickness $B$ is readily calculated from the $J$-integral as follows:
\begin{equation} \label{eq:7}
G\left(\varphi\right)=\frac{1}{B} J_k z_k(\varphi) = \frac{1}{B} \left( J_1\cos\varphi+J_2\sin\varphi \right)\ .
\end{equation}
The coordinates $J_1$ and $J_2$ are interpreted in a local crack tip coordinate system $(\mathbf{e}_1,\mathbf{e}_2)$, see Fig.~\ref{Fig0}, just as the deflection angle $\varphi$.

The figure further illustrates the anisotropies of elastic coefficients, indicated by $E(\varphi^{\ast})$ and the dashed blue elliptic line, and of crack resistance $J_C(\varphi^{\ast})$, represented by the dashed red ellipse.
The axis of transversal anisotropy is determined by the predominant direction of fiber orientation (PD).
The crack resistance is largest in the transverse direction (TD), whereas the elastic stiffness adopts the largest magnitude along the fibers.
The angle $\varphi^{\ast}=\gamma_{\text{loc}}+\varphi$ describes the crack propagation $\mathbf{z}$ and the directional material properties from the perspective of the PD.
The angle-dependent spatial interpolations of the latter follow elliptic shapes, e.g.\ the crack growth resistance in terms of the critical $J$-integral is formulated as follows:
\begin{equation} \label{eq:8}
\frac{1}{\left(J_{C}(\varphi^{\ast})\right)^{n/2}} 
=\frac{\cos^2(\varphi^{\ast})}{\left(J_{C}^\text{PD}\right)^{n/2}} +\frac{\sin^2(\varphi^{\ast})}{\left(J_{C}^\text{TD}\right)^{n/2}}\quad , \quad n \in \mathbb{R}^+\ .
\end{equation}
For $n=2$ a classical ellipse is obtained, in \cite{judt1} $n\approx 4$ produced the best coincidence with experimental data of short fiber reinforced polypropylene.
The crack growth resistances $J_C=K_{IC}^2 /E^\prime$ in Eq.~(\ref{eq:8}) are related to the fracture toughnesses $K_{IC}$ in the two perpendicular directions PD and TD, whereupon $E^\prime=E$ for plane stress and $E^\prime=E/(1-\nu^2)$ for plane strain conditions, with Young's modulus $E$ and Poisson's ratio $\nu$.
The anisotropy ratio is now introduced as follows:
\begin{equation} \label{eq:9}
\chi = \frac{K_{IC}^\mathrm{TD}}{K_{IC}^\mathrm{PD}} = \sqrt{\frac{J_{C}^\mathrm{TD}}{J_{C}^\mathrm{PD}}} \ .
\end{equation}
Eq.~(\ref{eq:8}) is resolved for the crack growth resistance, yielding
\begin{equation} \label{eq:10}
J_C\left(\varphi\right)=\frac{J_C^\mathrm{PD}J_C^\mathrm{TD}}{\left[\left(J_C^\mathrm{TD}\right)^{n/2}\cos^2(\gamma_{\text{loc}}+\varphi) + \left(J_C^\mathrm{PD}\right)^{n/2}\sin^2(\gamma_{\text{loc}}+\varphi)\right]^{2/n}}\ .
\end{equation}
The anisotropic deflection criterion postulates that a crack  seeks an optimum compromise of taking the path of least resistance on the one hand, and reducing the total potential energy of the structure by maximizing the energy release rate on the other \cite{judt2}.
Accordingly, Eqs.~(\ref{eq:7}) and (\ref{eq:10}) have to be merged in a quotient
\begin{equation} \label{eq:11}
\begin{array}{l} \displaystyle
J_Q(\varphi)=\frac{G(\varphi)}{J_C(\varphi)} = \frac{\left[\left(J_C^\mathrm{TD}\right)^{n/2}\cos^2(\gamma_{\text{loc}}+\varphi) + \left(J_C^\mathrm{PD}\right)^{n/2}\sin^2(\gamma_{\text{loc}}+\varphi)\right]^{2/n}}{J_C^\mathrm{PD}J_C^\mathrm{TD}} \cdot \\
\ \\
\hspace{75pt}\cdot \left( J_1\cos\varphi+J_2\sin\varphi \right) \ ,
\end{array}
\end{equation}
where a unit thickness $B=1$ has been inserted without loss of generality.
Following the above idea of the deflection criterion, the conditions
\begin{equation} \label{eq:12}
\frac{\partial J_Q (\varphi)}{\partial\varphi}=0\qquad\text{and}\qquad
\frac{\partial^2 J_Q (\varphi)}{\partial\varphi^2}<0
\end{equation}
have to hold for the deflection angle $\varphi$, accounting for both elastic and crack growth resistance anisotropies.
The critical condition of crack growth is further satisfied if $J_Q=1$. 
Introducing the crack tip loading ratio
\begin{equation} \label{eq:13}
\beta=\frac{J_2}{J_1}
\end{equation}
the equality in Eq.~(\ref{eq:12}) finally yields the following nonlinear algebraic equation for the deflection angle to be necessarily satisfied:
\begin{equation} \label{eq:14}
\begin{array}{l}
\displaystyle
\frac{4\left(1-\chi^n\right)}{n}\sin (\gamma_{\text{loc}}+\varphi) \cos (\gamma_{\text{loc}}+\varphi) \left(\cos\varphi + \beta\sin\varphi\right) \\
\ \\
+ \left(\beta\cos\varphi-\sin\varphi\right)\left(\chi^n \cos^2 (\gamma_{\text{loc}}+\varphi)+\sin^2 (\gamma_{\text{loc}}+\varphi)\right)=0
\end{array} \ .
\end{equation}
Both real and complex values for the deflection angle are obtained, depending on the anisotropy ratio $\chi$ and the interpolation parameter $n$.
The inequality of Eq.~(\ref{eq:12}) must be satisfied in order to eventually determine possible deflection angles.
The limiting case of isotropic crack growth resistance is obtained for $\chi\to 1$, whereupon the first summand of Eq.~(\ref{eq:14}) is cancelled out, leaving
\begin{equation} \label{eq:15}
\left(\beta\cos\varphi-\sin\varphi\right)\left(\cos^2 (\gamma_{\text{loc}}+\varphi)+\sin^2 (\gamma_{\text{loc}}+\varphi)\right)=\beta\cos\varphi-\sin\varphi=0 \ .
\end{equation}
It is obvious that this condition is uniquely satisfied if
\begin{equation} \label{eq:16}
\varphi=\arctan\beta=\arctan\left(\frac{J_2}{J_1}\right) \ ,
\end{equation}
which is the classical $J$-integral criterion of crack deflection \cite{ma05}.

The elastic anisotropy is often negligible compared to the directional dependence of the fracture toughness.
For an elastically isotropic model, the relations \cite{bergez}
\begin{equation} \label{eq:17}
J_1=\frac{K_I^2+K_{II}^2}{E'}\ ,\qquad J_2=-\frac{2K_I K_{II}}{E'}
\end{equation}
and
\begin{equation} \label{eq:18}
\beta=-\frac{2\tan\Phi}{1+\tan^2\Phi} = -\sin (2\Phi) \ ,
\end{equation}
where the mixed-mode ratio $\Phi$ from Eq.~(\ref{eq:1}) has been inserted, provide a condition for the deflection angle being equivalent to Eq.~(\ref{eq:14}), i.e.,
\begin{equation} \label{eq:19}
\begin{array}{l}
\displaystyle
\frac{4\left(1-\chi^n\right)}{n}\sin (\gamma_{\text{loc}}+\varphi) \cos (\gamma_{\text{loc}}+\varphi) \left(\cos\varphi -\sin (2\Phi)\sin\varphi\right) \\
\ \\
- \left(\sin (2\Phi)\cos\varphi+\sin\varphi\right)\left(\chi^n \cos^2 (\gamma_{\text{loc}}+\varphi)+\sin^2 (\gamma_{\text{loc}}+\varphi)\right)=0
\end{array} \ ,
\end{equation}
Assuming pure mode-I loading, i.e. $\beta=\Phi=0$, Eqs.~(\ref{eq:14}) and (\ref{eq:19}) likewise simplify as follows:
\begin{equation} \label{eq:20}
\begin{array}{l}
\displaystyle
\frac{4\left(1-\chi^n\right)}{n}\sin (\gamma_{\text{loc}}+\varphi) \cos (\gamma_{\text{loc}}+\varphi) \cos\varphi \\
\ \\
- \sin\varphi \left(\chi^n \cos^2 (\gamma_{\text{loc}}+\varphi)+\sin^2 (\gamma_{\text{loc}}+\varphi)\right)=0
\end{array} \ .
\end{equation}
If the fibers are uniformly aligned perpendicular to the crack, the angle of the PD thus being $\gamma_{\text{loc}}=\pi/2$, see Fig.~\ref{Fig0}, Eq.~(\ref{eq:20}) is reduced to
\begin{equation} \label{eq:21}
\left(
1 + \chi^n\tan^2\varphi + \frac{4\left(1-\chi^n\right)}{n}\right)
\sin\varphi = 0
\end{equation}
and possible deflection angles are obtained as 
\begin{equation} \label{eq:22}
\varphi=\pm\arctan\sqrt{\frac{4\left(\chi^n-1\right)-n}{n\,\chi^n}}
\qquad\mbox{and}\qquad
\varphi=0 \ .
\end{equation}
The square root produces real numbers and thus physical results if
\begin{equation} \label{eq:23}
\chi\geq\sqrt[\displaystyle n]{\frac{n}{4}+1} \ .
\end{equation}
Accordingly, a mode-I loaded crack perpendicular to the PD is deflected in either the positive or negative half space if a threshold value of the anisotropy ratio, depending on the interpolation coefficient $n$, is exceeded. 
In this case, the option $\varphi=0$ constitutes a minimum of $J_Q(\varphi)$ and is thus excluded.
If $\chi$ is below the threshold, $\varphi=0$ is the only real valued solution going along with a maximum of the function $J_Q(\varphi)$.

\subsection{Stochastic modeling} \label{ssec:stoch}

As elaborated in Sect.~\ref{ssec:crack} above, the crack deflection angle is significantly influenced, among others, by the PD of fiber alignment at the crack tip. 
Employing the global coordinate system $(\tilde{\mathbf{e}}_1,\tilde{\mathbf{e}}_2)$ and the corresponding orientation angle $\gamma=\gamma_{\text{loc}}-\alpha$ introduced in Fig.~\ref{Fig0}, the PD is represented by either of the unit vectors
\begin{equation}\label{eq:PD}
\pm(\cos(-\gamma)\tilde{\mathbf{e}}_1+\sin(-\gamma)\tilde{\mathbf{e}}_2)\ .
\end{equation}
In order to account for statistical variability of the PD among different specimens and among different locations within one and the same specimen, the orientation angle $\gamma=\gamma(\mathbf{x})$ at each position $\mathbf{x}\in\mathcal D$ is considered as a random variable. 
Here and below $\mathcal D$ denotes the two-dimensional domain representing the specimen. 

There exists a substantial literature on circular data and a large variety of circular models such as von Mises distributions, projected normal distributions or wrapped distributions, to name but a few, see e.g.\ \cite{fisher, jammalamadaka, ley, mardia}.  
In the present study, the orientation angle $\gamma$ is modeled as a Gaussian random variable with mean $\mu_\gamma$ and standard deviation $\sigma_\gamma$. 
Due to $\pi$-periodicity in \eqref{eq:PD}, it is clear that $\gamma$ can be replaced by the wrapped variable $(\gamma+\pi/2)(\text{mod}\,\pi)-\pi/2$ taking values in $[-\pi/2,\pi/2)$.  
By slight abuse of notation, the latter random variable, sometimes referred to as an axial variable in the literature, will also be denoted by $\gamma$. 
It follows a modified wrapped normal distribution with probability density function 
on $[-\pi/2,\pi/2)$ given by
\begin{equation}\label{eq:wrappedGaussian}
f_\gamma(\theta) = \frac{1}{\sqrt{2\pi}\,\sigma_\gamma}\sum_{k\in\mathbb Z}\exp\!\left(-\frac{(\theta+k\pi-\mu_\gamma)^2}{2\sigma_\gamma^2}\right)\ .
%,\quad x\in\left(-\frac{\pi}{2},\frac{\pi}{2}\right].
\end{equation}
This model choice offers the advantage of being compatible with Gaussian random field models \cite{jona-lasinio}, a feature it has in common with projected normal distributions \cite{wang}. 
Moreover, it is known that von Mises distributions, which play a core role among circular distributions, can in certain parameter ranges be approximated by wrapped normal distributions \cite{mardia}. 
%, cf.~\cite[Sect. 3.5]{mardia}.
When it comes to identifying the distribution parameters of $\gamma$, it is important to note that $\mu_\gamma$ and $\sigma_\gamma$ do not coincide with the usual mean and standard devitation associated to the wrapped density function in Eq.~\eqref{eq:wrappedGaussian}. 
Instead, the distribution parameters are represented in terms of suitable modifications of the concepts of mean direction and circular standard deviation from directional statistics \cite{mardia}  
%\cite[Sect.~3.4]{mardia} 
for axial distributions, given by 
\begin{equation}\label{eq:musigma}
\mu_\gamma = \frac{1}{2}\arg(\mathbb{E}(e^{i2\gamma}))
\qquad\text{and}\qquad
\sigma_\gamma = \sqrt{-\frac{1}{2}\ln(|\mathbb{E}(e^{i2\gamma})|)}\ ,
%\mu_\gamma = \frac{1}{2}\arg(\mathbb{E}[e^{i2\gamma}])
%\qquad\text{and}\qquad
%\sigma_\gamma = \sqrt{-\frac{1}{2}\ln(|\mathbb{E}[e^{i2\gamma}]|)}\ ,
%%\sigma_\gamma^2 = -\frac{1}{2}\ln(|\mathbb{E}[e^{i2\gamma}]|) ,
\end{equation}
where 
$\mathbb E$ denotes the expectation, i.e.
\begin{equation}
\mathbb{E}(e^{i2\gamma})=\int_{-\pi/2}^{\pi/2}e^{i2\theta}f_\gamma(\theta)\,\mathrm{d}\theta,
\end{equation} 
and for definiteness the argument of the complex number $\mathbb E(e^{i2\gamma})$ is taken as a value in $[-\pi,\pi)$. 
The identities in Eq.~\eqref{eq:musigma} are particularly useful for the construction of consistent estimators for $\mu_\gamma$ and $\sigma_\gamma$. 

Up to this point, the distribution of $\gamma=\gamma(\mathbf x)$ has been defined for single positions $\mathbf x\in\mathcal D$ according to Eq.~\eqref{eq:wrappedGaussian} with $\mu_\gamma=\mu_{\gamma}(\mathbf x)$ and $\sigma_\gamma=\sigma_{\gamma}(\mathbf x)$. 
To specify a random field $(\gamma(\mathbf x))_{\mathbf x\in\mathcal D}$ and thus the spatial variability of the PD within the specimen, it is necessary to additionally prescribe the joint distribution of any finite family $\gamma(\mathbf x_1),\gamma(\mathbf x_2),\ldots,\gamma(\mathbf x_n)$. 
In this context, it is convenient to interpret each $\gamma(\mathbf x)$ as an ordinary Gaussian random variable rather than a wrapped variable, implying that $\mathbb E(\gamma(\mathbf x))=\mu_{\gamma}(\mathbf x)$ and $\text{Var}(\gamma(\mathbf x))=\sigma_{\gamma}^2(\mathbf x)$. 
The family $(\gamma(\mathbf x))_{\mathbf x\in\mathcal D}$ is further assumed to be a Gaussian random field, so that all joint distributions are uniquely determined by the mean function $\mathbf x\mapsto \mathbb E(\gamma(\mathbf x))$ and the covariance function $(\mathbf x,\mathbf x')\mapsto\mathrm{Cov}(\gamma(\mathbf{x}),\gamma(\mathbf{x}'))$. 
The former models the expected PD of fiber alignment parallel to the flow front of the injection molding process, while the latter encodes the correlation structure of stochastic fluctuations of the PD. 
Overviews and discussions of common covariance models for random fields can be found e.g.\ in \cite{matern, rasmussen, stein}, a particularly popular and flexible class being given by the Whittle-Mat\'ern covariance functions 
\begin{equation}\label{eq:matern}
\mathrm{Cov}(\gamma(\mathbf{x}),\gamma(\mathbf{x}'))=
\sigma_{\gamma}(\mathbf{x})\sigma_{\gamma}(\mathbf{x}')
%\sigma_\gamma^2
\frac{2^{1-\nu}}{\Gamma(\nu)}\bigg(\frac{\sqrt{2\nu}\|\mathbf{x}-\mathbf{x}'\|}{\ell}\bigg)^\nu K_\nu\bigg(\frac{\sqrt{2\nu}\|\mathbf{x}-\mathbf{x}'\|}{\ell}\bigg)\ ,
\end{equation}
where $\|\cdot\|$ denotes the Euclidean norm and $\Gamma$, $K_\nu$ are the gamma function and the modified Bessel function of the second kind, respectively. 
The parameter $\nu>0$ is related to the smoothness of the random field, whereas $\ell>0$ defines the characteristic length scale and measures how quickly the correlations decay with distance. 
In the limit $\nu\to\infty$ the Wittle-Mat\'ern model leads to the squared exponential model 
\begin{equation}\label{eq:SE}
\mathrm{Cov}(\gamma(\mathbf{x}),\gamma(\mathbf{x}'))=
\sigma_{\gamma}(\mathbf{x})\sigma_{\gamma}(\mathbf{x}')
%\sigma_\gamma^2
\exp\!\left(-\frac{\|\mathbf{x}-\mathbf{x}'\|^2}{2\ell^2}\right).
\end{equation}
For the sake of simplicity, the present study is confined to covariance functions of type Eq.~\eqref{eq:SE}, although this model is known to have some undesirable features with respect to asymptotic properties of linear predictors \cite{stein}. 
As far as a more thorough statistical analysis is concerned, alternative covariance models such as the one in Eq.~\eqref{eq:matern} are certainly preferable. 
As a further simplification, the standard deviation $\sigma_{\gamma}(\mathbf x)$ is assumed to be constant and denoted by $\sigma_\gamma$, so that $\sigma_{\gamma}(\mathbf{x})\sigma_{\gamma}(\mathbf{x}')$ simplifies to $\sigma_\gamma^2$.
At this point it seems appropriate to also mention the classical Folgar-Tucker model for the distribution of fiber orientations in the context of injection molding, which, however, usually does not capture spatial correlations in the sense described above, see e.g.\ \cite{ospald} and the references therein.

While the focus of stochastic modeling in this article lies on the PD of fiber orientation, other sources of uncertainty influencing the crack propagation may be taken into account in a similar fashion.  
For instance, stochasticity of the anisotropy ratio $\chi$ in Eq.~\eqref{eq:9} is additionally presumed for some of the density approximations and crack path simulations presented in Sects.~\ref{subsec:deflect} and \ref{subsec:sim} below.

\section{Investigation of fiber parameters in the $\mu$--CT}
\label{sec:CT}

\begin{figure}[!ht]
\begin{centering}
\includegraphics[width=.95\textwidth]{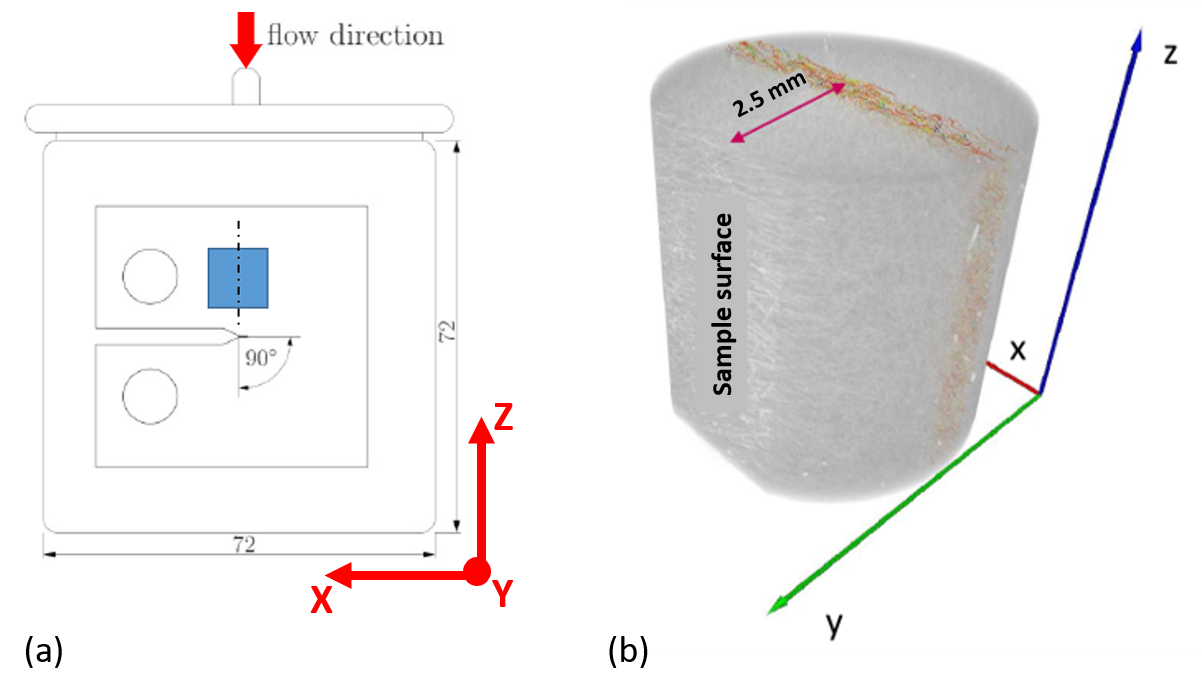}
\par\end{centering}
\protect\caption{(a) CT-specimen with indicated flow direction of injection molding, blue square: domain investigated in the $\mu$--CT; (b) recording volume and fiber layer (thickness 0.6mm)}
\label{Fig8}
\end{figure}

Fiber orientations in the CT-specimens employed for crack growth experiments have been investigated in the $\mu$--CT, 
in order to obtain an idea of reasonable values of standard deviations as a basis for the numerical simulation of crack paths \cite{kahl18,zarg20}.  
Fig.~\ref{Fig8} (a) depicts the specimen and its dimensions and indicates the flow direction of the injection molding process. 
The flow front thus roughly follows the shape of an upward opening parabola with axis of symmetry along the vertical dashed line segment \cite{judt2}. 
The blue square right above the notch tip indicates the part of the specimen which was investigated in the $\mu$--CT, containing the apex of the flow front. 
The distribution parameters of the PD of fibers, denoted by $\mu_\gamma$ and $\sigma_\gamma$ in Eqs.~\eqref{eq:wrappedGaussian} and \eqref{eq:musigma}, are approximately constant within this region. 
In particular, the expected PD is perpendicular to the $z$-axis and along the notch, i.e.\ $\mu_\gamma=0^{\circ}$.

In Fig.~\ref{Fig8} (b) the grey volume represents the domain of investigation, beginning at one of the large surfaces of the sample. 
Its diameter spans half of the specimen, the latter exhibiting a thickness of 10mm.
A layer of fibers with a thickness of 0.6mm, recorded around the center of the grey domain, is visible at its surface.
The dominating red color confirms the expected predominant direction, whereas fibers in blue are aligned perpendicular to the notch.

Theoretically, consistent estimators for the distribution parameters $\mu_\gamma$ and $\sigma_\gamma$ based on the identities in Eq.~\eqref{eq:musigma} are given by
\begin{equation}\label{eq:estimators}
\bar{\gamma}
=\frac{1}{2}\arg\!\bigg(\frac{1}{N}\sum_{k=1}^{N}e^{i2\gamma_k}\bigg)
\qquad\text{and}\qquad s_\gamma=\sqrt{-\frac{1}{2}\ln\!\bigg(\bigg|\frac{1}{N}\sum_{k=1}^{N}e^{i2\gamma_k}\bigg|\bigg)},
\end{equation}
where $\gamma_1,\gamma_2,\ldots,\gamma_N$ is a sample from the distribution of $\gamma$ 
such that the orientation vectors $e^{i2\gamma_1},e^{i2\gamma_2},\ldots,e^{i2\gamma_N}$  satisfy the law of large numbers as $N\to\infty$. 
To ensure consistency of $\bar\gamma$ in the case $\mu_\gamma=-\pi/2$, the range $[-\pi/2,\pi/2)$ of possible values is equipped with the topology resulting from identifying $-\pi/2$ with $\pi/2$. 
However, since $\gamma$ models the PD of fiber orientations within the fracture process zone, the typical size of the latter needs to be specified in order to be able to calculate the sample values $\gamma_k$ as averages of fiber orientations in suitable volume elements. 
This lies beyond the scope of the present study. 
The investigation is therefore focused on the statistical properties of the orientations of single fibers instead of volume-averaged fiber orientations. 
Assuming sufficient spatial ergodicity, estimates of the corresponding distribution parameters are obtained by replacing the sample values $\gamma_k$ in Eq.~\eqref{eq:estimators} with the 
values $\gamma_k^\text{sing}$ of single fiber orientations measured in the colored layer in Fig.~\ref{Fig8}~(b). 
While the mean orientation $\bar\gamma^\text{sing}$ calculated this way constitutes a suitable estimate for $\mu_\gamma$, the sample circular standard deviation $s_\gamma^\text{sing}$ of the single fiber orientations can only serve as an estimated upper bound for
$\sigma_\gamma$, since volume averaging decreases variability.

\begin{figure}[!ht]
\centering
\subfigure[Positions and angles of fibers within the layer of Fig.~\ref{Fig8}]
%{\includegraphics[width=.7\textwidth]{Fig9b.png}}\vfill
{\includegraphics[width=.91\textwidth]{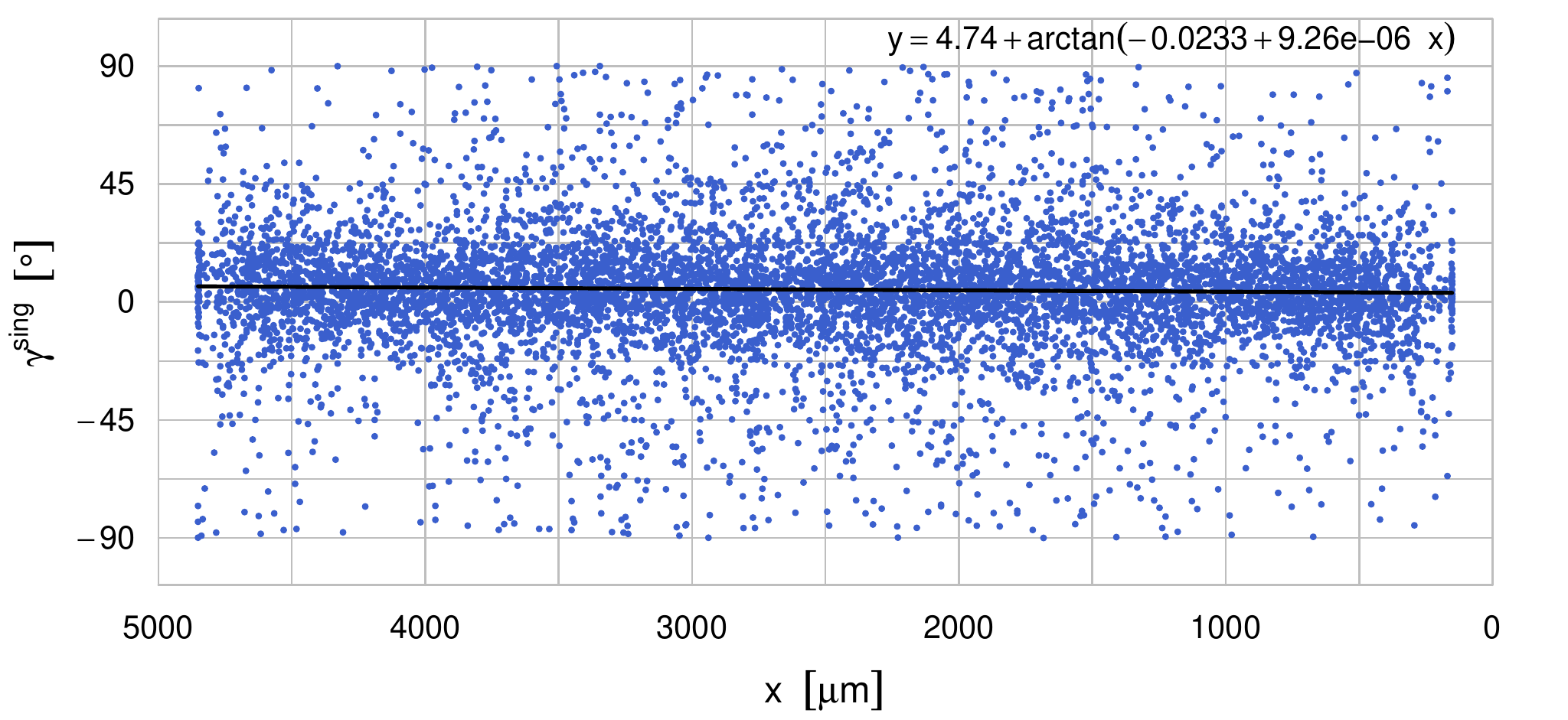}}\vfill
\subfigure[Fiber orientations yielding circular standard deviation $22.8^\circ$]
%{\includegraphics[width=.7\textwidth]{Fig9.png}}
{\includegraphics[width=.9\textwidth]{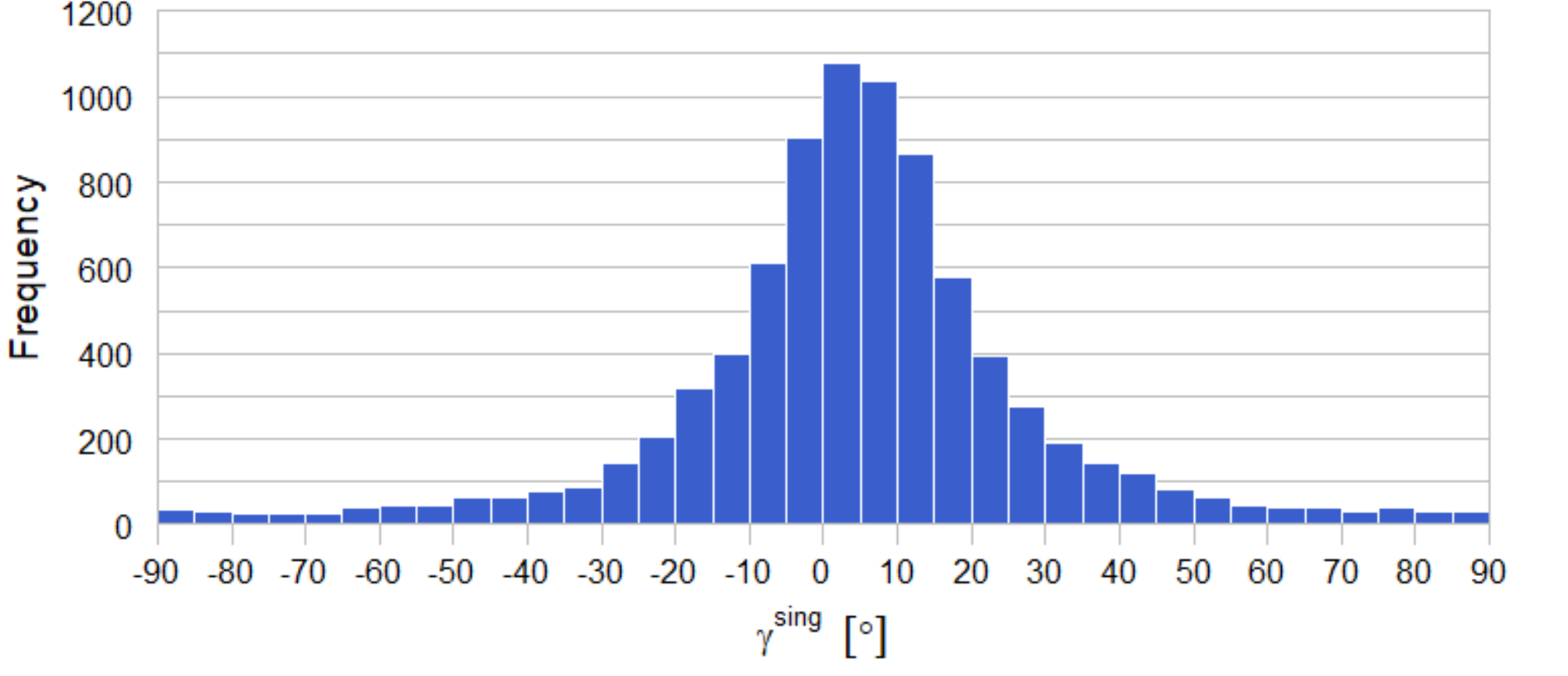}}
\caption{Distribution of 8224 fiber orientation angles from investigations in the $\mu$--CT according to Fig.~\ref{Fig8} for polypropylene matrix reinforced with 30 wt-\% regenerated cellulose fibers without coupling agent}
\label{Fig9}
\end{figure}

In Fig.~\ref{Fig9} (a) each dot represents one out of 8224 fibers and indicates its position along the $x$-axis according to Fig.~\ref{Fig8} and its orientation angle $\gamma^{\text{sing}}$, measured analogously to the angle of PD $\gamma=\gamma_{\text{loc}}$ within the context of Fig.~\ref{Fig0} ($\alpha=0$). 
A trend curve of the form $y=\mu+\arctan(\beta_0+\beta_1x)$ with parameters $\mu\in[-\pi/2,\pi/2)$, $\beta_0,\beta_1\in\mathbb R$ has been fitted to the data, however, there is no distinct local influence of the $x$-variable within the investigated range of $5$mm. 
The considered class of regression functions is directly suggested by the parabolic shape of the flow front, as a straightforward calculation shows. 
It is further worth noting that, since standard least squares regression is not appropriate for circular or axial response variables, 
%cf.~\cite[Sect.~11.3]{mardia}, 
a suitable modification of least circular distance regression based on maximizing the expression 
\begin{align}\label{eq:regression}
\sum_{k=1}^{8224}
\cos\big[2\big(\gamma^{\text{sing}}_k-\mu-\arctan(\beta_0+\beta_1 x_k)\big)\big]
\end{align}
has been used, cf.~\cite{lund}, where $\gamma^{\text{sing}}_k$ and $x_k$ and  are the observed orientation angles and positions, respectively. 
The maximization of \eqref{eq:regression} can be carried out by means of the optimization algorithm proposed in \cite[Sect.~3.2]{fisher2}, which has originally been formulated in the context of von Mises distribution assumptions but is also applicable in the present situation; it is implemented, e.g., in the R package \lq circular\rq\ \cite{circular}. 
Fig.~\ref{Fig9} (b) shows the fiber orientations in a histogram providing 
a mean orientation $\bar\gamma^\text{sing}=4.7^{\circ}$ and a sample circular standard deviation $s_{\gamma}^\text{sing}=22.8^{\circ}$. 

For the lengths of the fibers an average of 1141$\mu$m and a standard deviation of 1136$\mu$m have been identified. 
A value $\ell=2.5$mm of magnitude slightly larger than the average fiber length is used as characteristic length scale, see Eq.~\eqref{eq:SE}, in the crack path simulations in Sect.~\ref{subsec:sim} below.
For similar reasons as discussed with regard to the parameters $\mu_\gamma$ and $\sigma_\gamma$ above, a rigorous estimation of the correlation parameter $\ell$ requires a detailed analysis of the typical size of the fracture process zone and is not treated in this study.

\section{Predictions of crack deflection and crack paths}
\label{sec:results}
\subsection{Crack deflection under mode-I loading} \label{subsec:deflect}

Combining the characterization of the deflection angle $\varphi$ as a maximizer of the quotient of energy release rate and crack growth resistance in Eq.~\eqref{eq:11} with the modeling of the angle of PD of fiber alignment $\gamma$ as a random variable in Eq.~\eqref{eq:wrappedGaussian} implies that $\varphi$ has to be interpreted as a random variable too. 
More specifically, in view of the condition for $\varphi$ 
%in the case of negligible elastic anisotropy 
in Eq.~\eqref{eq:19} and the identity $\gamma_{\text{loc}}=\gamma+\alpha$, see Fig.~\ref{Fig0}, the crack deflection angle
\begin{align}
\varphi=\varphi(\gamma,\chi,\Phi,n,\alpha)
\end{align}
is implicitly given as a function of stochastic fiber orientation, fracture toughness anisotropy ratio, mixed-mode ratio, interpolation parameter of directional anisotropy,  
and current crack tip orientation. 
In the present subsection, the probability distribution of $\varphi$ and its dependence on the distribution parameters of $\gamma$ and the anisotropy ratio $\chi$, introduced in Eq.~\eqref{eq:9}, are investigated in the case of mode-I loading at the crack tip, i.e.\ $\Phi=0$, in which the condition in Eq.~\eqref{eq:19} simplifies to Eq.~\eqref{eq:20}. It is further assumed that  $n=4$, $\alpha=0^\circ$.

\begin{figure}[!ht]
\centering
\subfigure[Fiber parameters $\mu_\gamma=0^\circ,\,50^\circ,\,90^\circ$ (from left to right) and $\sigma_\gamma=10^\circ$]
{\includegraphics[width=\textwidth]{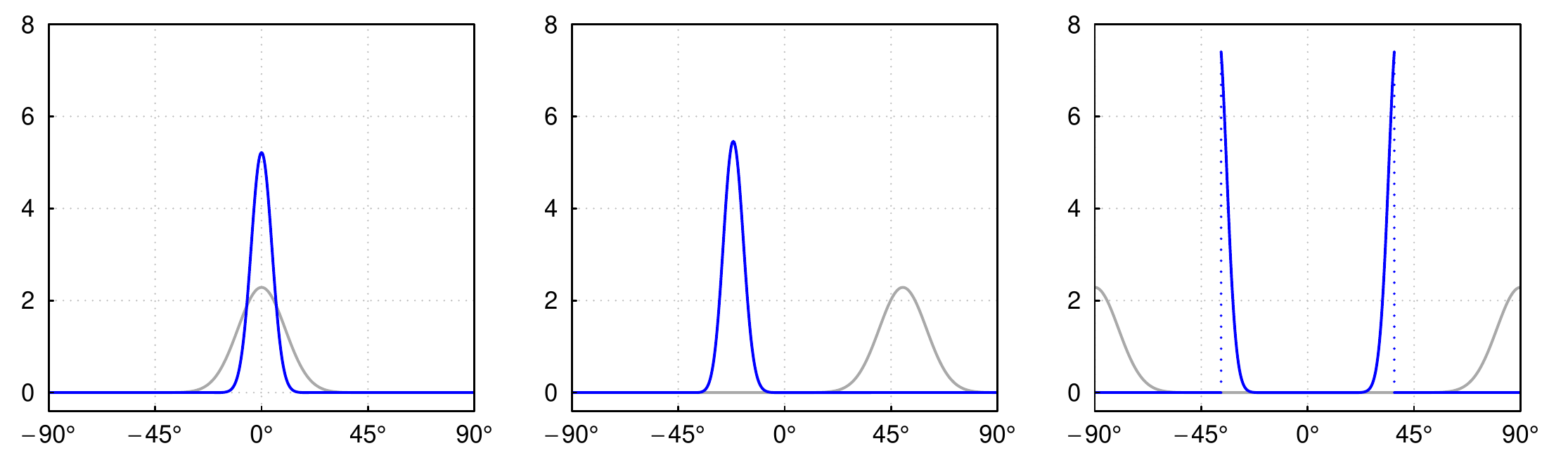}}\vfill
\subfigure[Fiber parameters $\mu_\gamma=0^\circ,\,50^\circ,\,90^\circ$ (from left to right) and $\sigma_\gamma=20^\circ$]
{\includegraphics[width=1\textwidth]{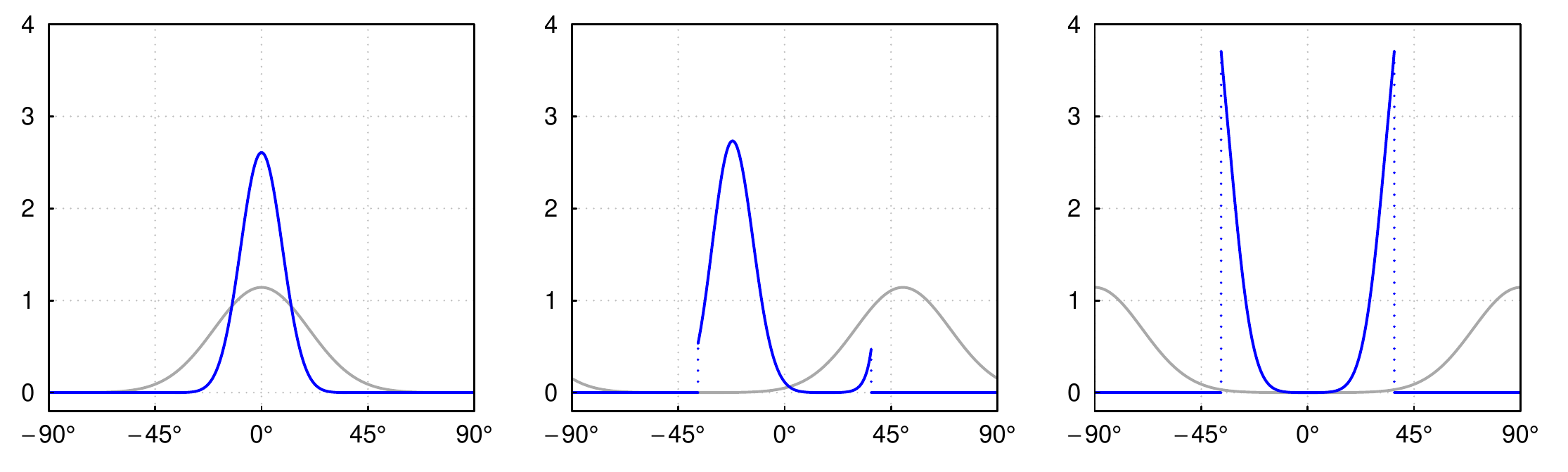}}
\caption{Probability density functions of deflection angle $\varphi$ (blue) and fiber orientation angle $\gamma=\gamma_\text{loc}$ (grey) for mode-I loading and different mean orientations $\mu_\gamma$ and standard deviations $\sigma_\gamma$; anisotropy ratio $\chi=1.46$, interpolation parameter $n=4$, crack tip orientation angle $\alpha=0^\circ$ (Fig.~\ref{Fig0})}
\label{FigDensity1}
\end{figure}

Fig.~\ref{FigDensity1} shows the graphs of the probability density functions $f_\varphi$ and $f_\gamma$ of $\varphi$ and $\gamma$ for different values of the distribution parameters $\mu_\gamma$ and $\sigma_\gamma$, with anisotropy ratio $\chi=1.46$ motivated by experimental data \cite{judt1}, see Sect.~\ref{subsec:exp}. 
As the mean orientation $\mu_\gamma$ increases from left to right, the probability mass of $\varphi$ is at first shifted in the opposite direction, reflecting the fact that the crack is seeking a direction of low material resistance. 
The range of $\varphi$ is limited by
\begin{align}\label{eq:phi_min_max}
\varphi_\text{min}(\chi)=\inf_{-\frac\pi2\le \gamma <\frac\pi2}\varphi(\gamma,\chi)
\qquad\text{and}\qquad 
\varphi_\text{max}(\chi)=\sup_{-\frac\pi2\le \gamma <\frac\pi2}\varphi(\gamma,\chi)
\end{align}
with the specific values $\varphi_\text{min}(1.46)\approx -36.8^\circ$ and $\varphi_\text{max}(1.46)\approx 36.8^\circ$, resulting in discontinuities of $f_\varphi$ at these points. 
As a consequence of $\pi$-periodicity, it can be observed that shifting probability mass of $\gamma$ beyond $90^\circ$ leads to increased values of $f_\varphi$ near $\varphi_\text{max}(\chi)$. 
This behavior culminates in the case $\mu_\gamma=\pm90^\circ$ of an expected PD  perpendicular to the crack, which exhibits spikes of $f_\varphi$ at both $\varphi_\text{min}(\chi)$ and $\varphi_\text{max}(\chi)$. 
Comparing Fig.~\ref{FigDensity1}~(a) and Fig.~\ref{FigDensity1}~(b) illustrates that an increased dispersion parameter $\sigma_\gamma$ of $\gamma$ entails an increased variability of $\varphi$. 

Here and below, the probability density function $f_\varphi$ is numerically approximated by means of a histogram-based approach using evaluations of $\varphi$ at $q$-quantiles of $\gamma$ for $q\in\{1/N,\,2/N,\ldots,1-1/N\}$, with $N\in\mathbb N$ chosen sufficiently large. 
For the considered values of $\sigma_\gamma$ and $N$, each $q$-quantile of $\gamma$ can be replaced by $\mu_\gamma+\sigma_\gamma z_q$, where $z_q$ denotes the $q$-quantile of the standard normal distribution. 
The employed density approximation thus amounts to
\begin{align}\label{eq:densityapprox}
f_\varphi(\theta)\approx\sum_{j=1}^M\mathbbm{1}_{I_j}(\theta)\,\frac{\#\big\{k\in\{1,\ldots,N-1\}:\varphi\big(\mu_\gamma+\sigma_\gamma z_{\frac{k}{N}},\,\chi\big)\in I_j\big\}}N,
\end{align}
where the intervals $I_j$ form an equidistant partition of $[\varphi_\text{min}(\chi),\varphi_\text{max}(\chi)]$, $M\in\mathbb N$ is the suitably chosen number of bins, $\mathbbm{1}_{I_j}$ is the indicator function of $I_j$, and $\#\{\ldots\}$ denotes the cardinality of a set. 
The step function in Eq.~\eqref{eq:densityapprox} is further modified into a continuous function on $[\varphi_\text{min}(\chi),\varphi_\text{max}(\chi)]$.
Since this purely deterministic approximation relies on unproven regularity properties of the transformation $\gamma\mapsto\varphi(\gamma,\chi)$, it has been numerically validated by means of a standard Monte Carlo method in which the quantiles of $\gamma$ are replaced by independent random realizations of $\gamma$, resulting in convergence to the same probability density function but with a slower rate.

\begin{figure}[!ht]
\begin{centering}
\includegraphics[width=.55\textwidth]{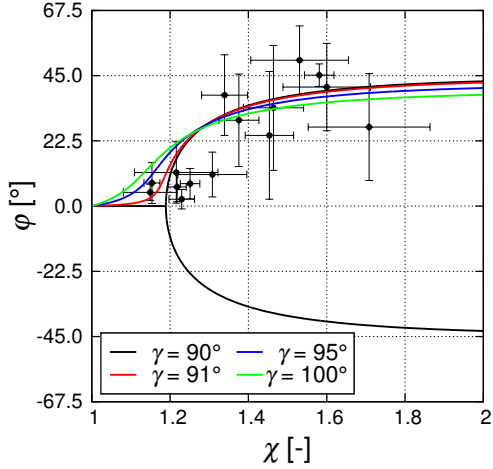}
\par\end{centering}
\protect\caption{Deflection angle versus anisotropy ratio for mode-I loading of a crack with deterministic fiber orientations of four different angles $\gamma=\gamma_{\text{loc}}$ (Fig.~\ref{Fig0}); the interpolation parameter of directional anisotropy is $n=4$ and crosses with centered dots depict ranges of experimental data \cite{judt1}}
\label{Fig1}
\end{figure}

In Fig.~\ref{Fig1} the crack deflection angle for mode-I loading, predicted by the $J$-integral criterion according to Eq.~(\ref{eq:20}), is plotted versus the anisotropy ratio $\chi$.
Different deterministic fiber orientations in a range of $\gamma=\gamma_{\text{loc}}=90^{\circ}$ to $100^{\circ}$ ($\alpha=0^{\circ}$), see Fig.~\ref{Fig0}, have been chosen to demonstrate their effect on an initial crack deflection, and in particular its dependence on the anisotropy ratio.
The crosses with the centered dots depict error bars of experimental results of CT-specimens made of RCF and GF reinforced PP \cite{judt2}.
The solid black line, representing a perpendicular fiber orientation with $\gamma=90^{\circ}$, exhibits an essential feature of anisotropic fracture toughness in terms of a bifurcation of deflection angles, emanating from Eqs.~(\ref{eq:22}) and (\ref{eq:23}).
Hence, for an interpolation factor $n=4$, a mode-I crack deflects above anisotropy ratios of $\chi=1.189$, and grows straight on below this threshold value.
Fig.~\ref{Fig1} illustrates, however, that even a small deviation from the perpendicular fiber orientation annihilates the bifurcation, leading to a smooth monotonous increase of deflection with increasing anisotropy ratios, instead.
Accordingly, a point of bifurcation is not expected either 
in the case of stochastic fiber orientations,  
which is supported by the experimental data.
For large anisotropy ratios, the relative impact on the crack deflection is distinctly reduced and the limiting value $\varphi_{\text{max}}=\arctan{(2/\sqrt{n})}=45^\circ$ is finally obtained from Eq.~(\ref{eq:22}) for $\chi\to\infty$.

\begin{figure}[!ht]
\centering
\includegraphics[width=\textwidth]{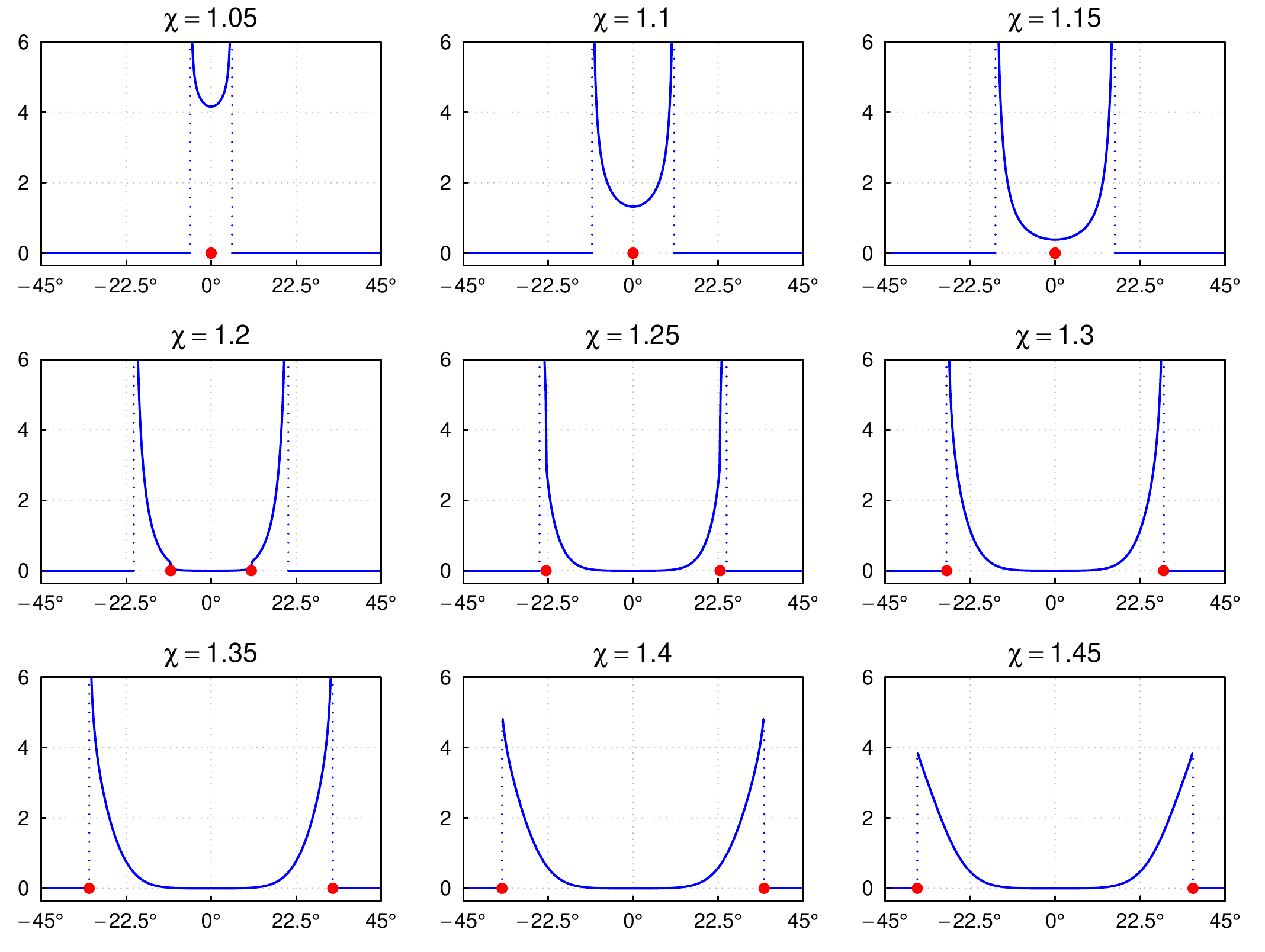}
\caption{
Probability density function of deflection angle $\varphi$ for mode-I loading, fiber parameters $\mu_\gamma=90^\circ$ and $\sigma_\gamma=20^\circ$, and different anisotropy ratios $\chi$; 
the red dots indicate the deflection angles in the case of deterministic fiber orientation $\gamma=\gamma_\text{loc}=90^\circ$;
interpolation parameter $n=4$, crack tip orientation angle $\alpha=0^\circ$ (Fig.~\ref{Fig0})
}
\label{FigDensity2}
\end{figure}

\begin{figure}[!ht]
\begin{centering}
\includegraphics[width=.58\textwidth]{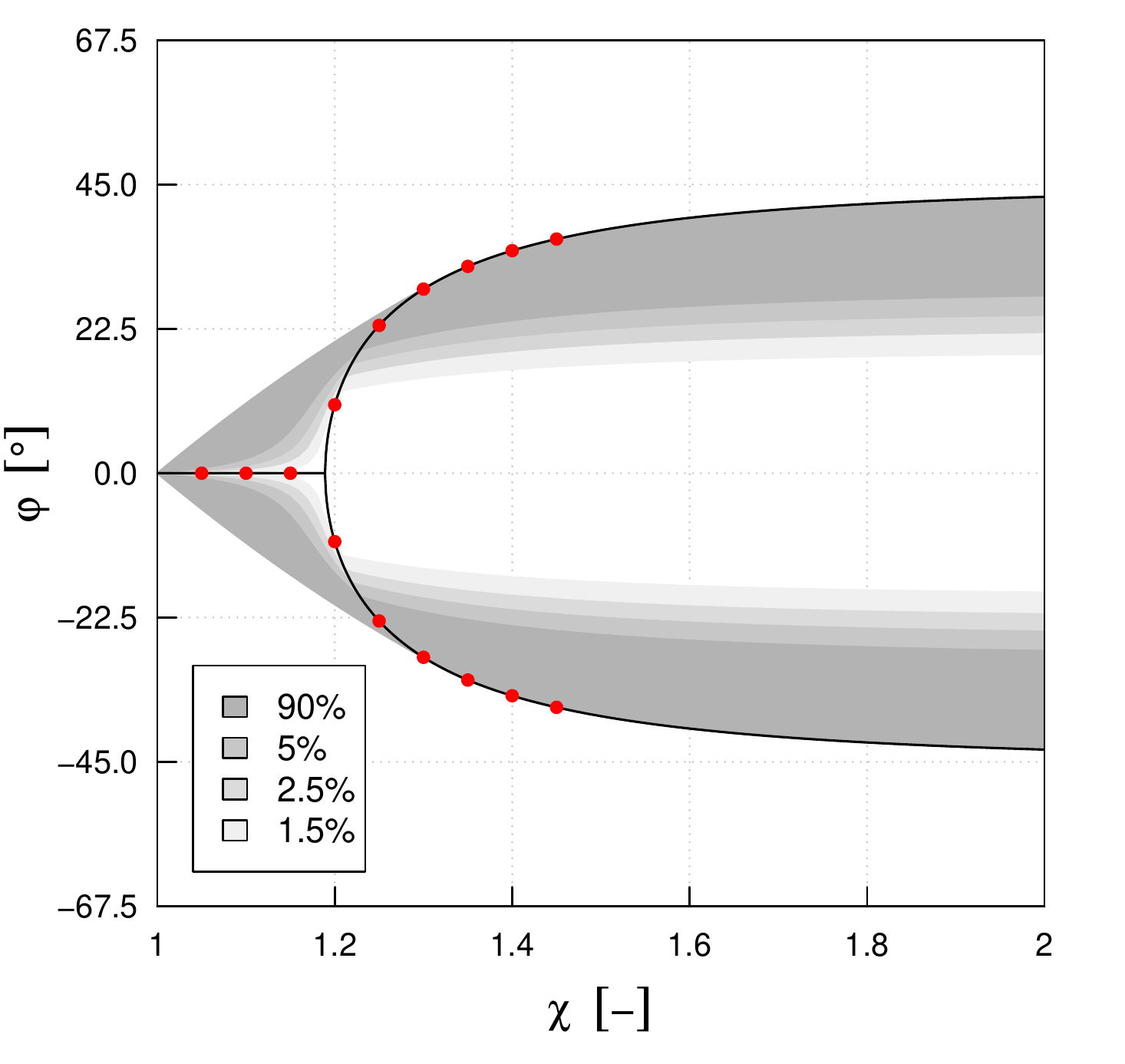}
\par\end{centering}
\protect\caption{
Bifurcation diagram in the style of Fig.~\ref{Fig1} for mode-I loading of a crack with deterministic fiber orientation $\gamma=\gamma_\text{loc}=90^\circ$ (black line) and prediction regions for the deflection angle in the case of stochastic fiber orientation with parameters $\mu_\gamma=90^\circ$ and $\sigma_\gamma=20^\circ$ (grey); 
interpolation parameter $n=4$, crack tip orientation angle $\alpha=0^\circ$ (Fig.~\ref{Fig0})
}
\label{FigBifurcationStoch}
\end{figure}

The influence of the anisotropy ratio $\chi$ on the deflection angle $\varphi$, depicted for deterministic fiber orientations in Fig.~\ref{Fig1}, is illustrated in the case of stochastic fiber orientations in Figs.~\ref{FigDensity2} and \ref{FigBifurcationStoch}.   
In Fig.~\ref{FigDensity2} the density of $\varphi$ is plotted for an expected PD of fiber alignment $\mu_\gamma=90^\circ$, dispersion parameter $\sigma_\gamma=20^\circ$, and various values of the anisotropy ratio $\chi$. 
For all choices of $\chi$, the density is symmetric and attains maximal values at the boundary points $\varphi_\text{min}(\chi)$ and $\varphi_\text{max}(\chi)$ introduced in Eq.~\eqref{eq:phi_min_max}, indicated by the vertical dotted lines. 
The latter drift apart continuously as $\chi$ increases, allowing for a wider spread of the probability mass, which is initially concentrated near $0^\circ$.  
The situation in the final plot ($\chi=1.45$) is similar to the one shown on the right hand side of Fig.~\ref{FigDensity1}~(b) ($\chi=1.46$).
For comparison purposes red dots have been added to indicate the corresponding values of $\varphi$ in the deterministic case $\gamma=90^\circ$, exhibiting a point of bifurcation between $\chi=1.15$ and $\chi=1.2$ as shown previously in Fig.~\ref{Fig1}.
The same values are also indicated in Fig.~\ref{FigBifurcationStoch}, where the bifurcation diagram from Fig.~\ref{Fig1} is enhanced by prediction regions for the stochastic deflection angle $\varphi$, specified by suitable quantiles of its probability distribution. 
For any fixed choice of $\chi$, the associated deflection angle attains a value in the dark grey region with probability $90\%$. This probability is increased to $95\%$, $97.5\%$, $99\%$, if the brighter shaded regions are added consecutively.
The upper and lower boundaries of the prediction regions are specified by the values $\varphi_\text{min}(\chi)$ and $\varphi_\text{max}(\chi)$, see Eq.~\eqref{eq:phi_min_max}.

\begin{figure}[!ht]
\centering
\subfigure[Fiber parameters $\mu_\gamma=0^\circ,\,50^\circ,\,90^\circ$ (from left to right) and $\sigma_\gamma=10^\circ$]
{\includegraphics[width=\textwidth]{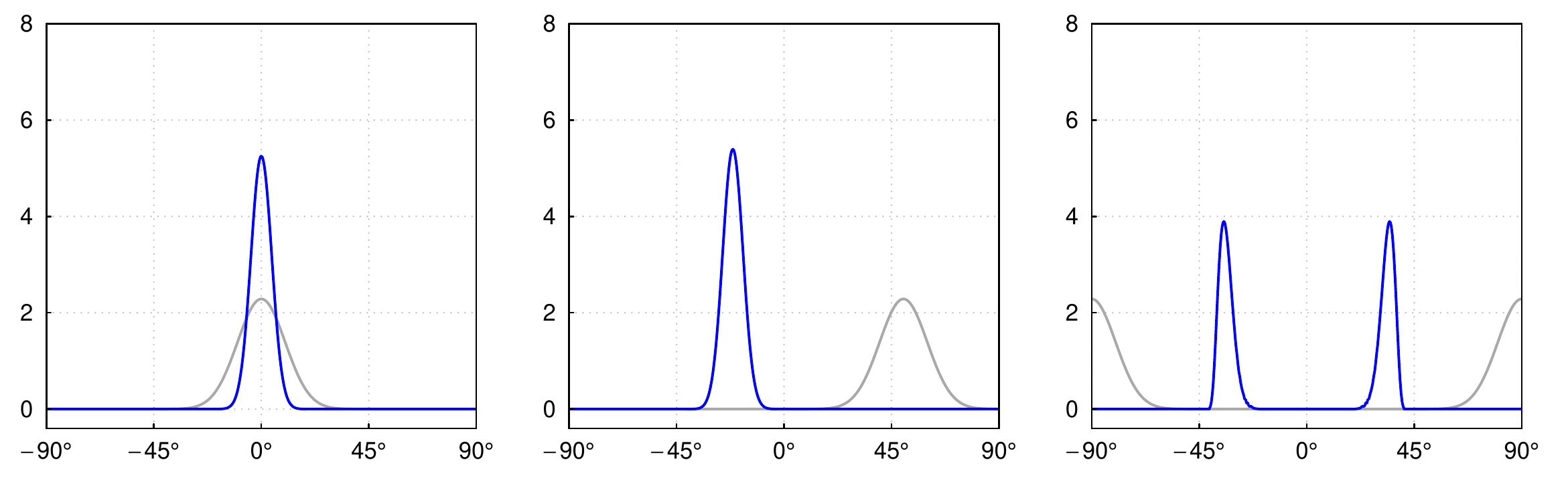}}\vfill
\subfigure[Fiber parameters $\mu_\gamma=0^\circ,\,50^\circ,\,90^\circ$ (from left to right) and $\sigma_\gamma=20^\circ$]
{\includegraphics[width=1\textwidth]{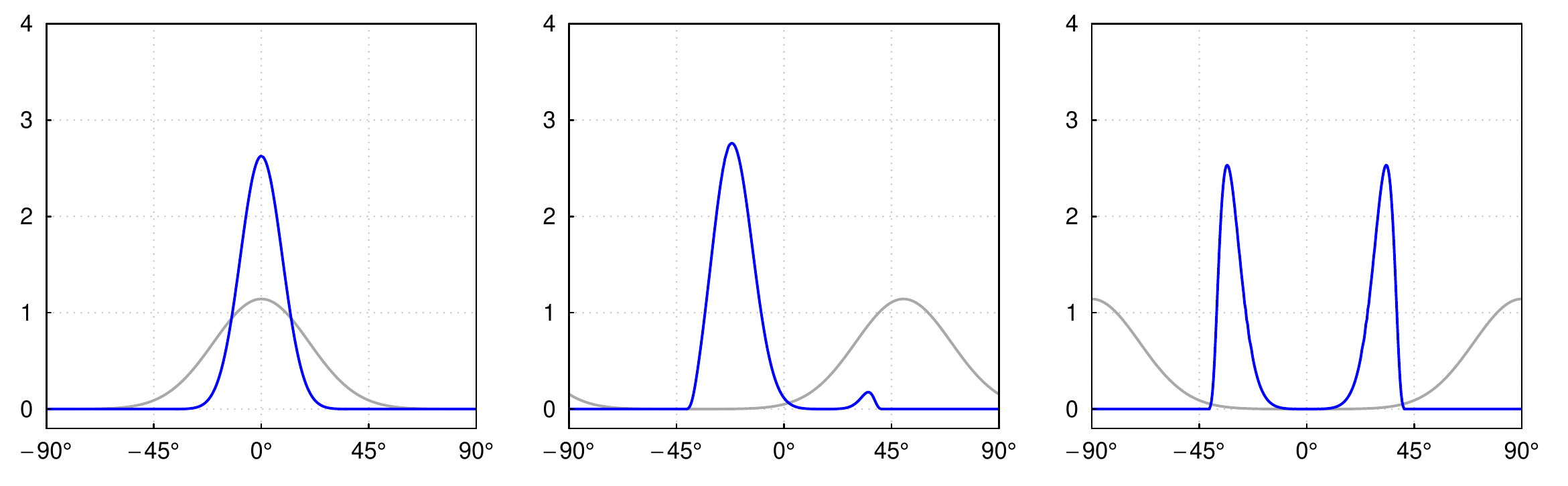}}
\caption{
Probability density functions of deflection angle $\varphi$ (blue) and fiber orientation angle $\gamma=\gamma_\text{loc}$ (grey) for mode-I loading, different fiber parameters, and stochastic anisotropy ratio with mean $\mu_\chi=1.46$ and standard deviation $\sigma_\chi=0.077$; interpolation parameter $n=4$, crack tip orientation angle $\alpha=0^\circ$ (Fig.~\ref{Fig0})
}
\label{FigDensity3}
\end{figure}

A comparison of the prediction regions in Fig.~\ref{FigBifurcationStoch} with the experimental data in Fig.~\ref{Fig1} suggests that further sources of uncertainty have to be taken into account in order to obtain a better match. 
While this is not the focus of the present study, a first step into this direction is presented in Fig.~\ref{FigDensity3}, where the density of $\varphi$ is plotted under the same assumptions as in Fig.~\ref{FigDensity1}, except that not only the fiber orientation $\gamma$ but also the anisotropy ratio $\chi$ is modeled as a random variable. 
Using parameters $\mu_\chi=1.46$ and $\sigma_\chi=0.077$ based on measurement data  \cite{judt1}, see Sect.~\ref{subsec:exp}, a normal distribution is assumed for simplicity, justified by the fact that the probability of unphysical realizations of $\chi$ less than one is thus negligibly small. 
Making further the simplified assumption of statistical independence of $\gamma$ and $\chi$, the density of $\varphi=\varphi(\gamma,\chi)$ is obtained by suitably averaging over the corresponding densities of $\varphi$ for deterministic $\chi$, 
resulting in a smoothing effect as opposed to Fig.~\ref{FigDensity1}.

\subsection{Crack paths from experiments} \label{subsec:exp}

\begin{figure}[!ht]
\begin{centering}
\includegraphics[width=.55\textwidth]{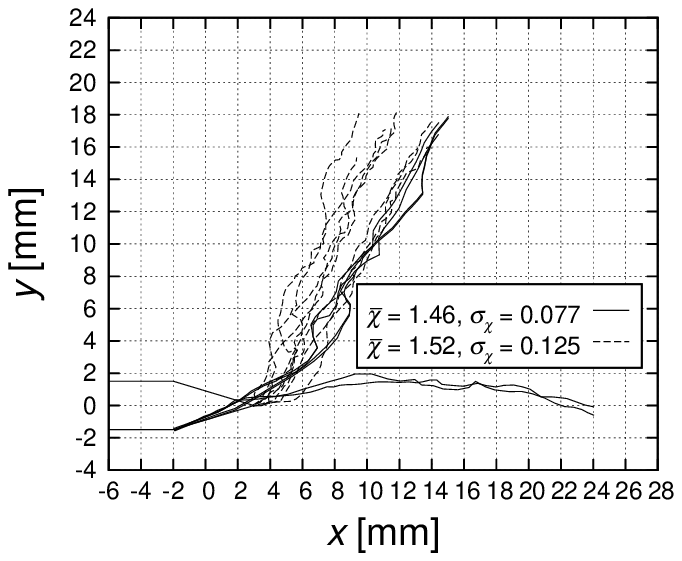}
\par\end{centering}
\protect\caption{Crack paths in CT-specimens with predominant fiber orientation perpendicular to the notch; two different compounds with and without coupling agent have been investigated, exhibiting average anisotropy ratios $\bar{\chi}=1.52$ and $\bar{\chi}=1.46$, respectively \cite{judt1}}
\label{Fig10}
\end{figure}

Fig.~\ref{Fig10} depicts various crack paths from experiments with basically the same CT-specimens providing the experimental ranges of initial deflection angles in Fig.~\ref{Fig1} \cite{judt1}.
The predominant fiber orientation PD is perpendicular to the notch.
Cracks deflecting in the negative $y$--half-plane have been reflected to the positive one in the plots.
While all samples consist of PP reinforced with 30 wt-\% RCF, the solid lines represent compounds without coupling agent (PP 30RCF), whereas the dashed lines are results of samples with 3 wt-\% maleated polypropylene as coupling agent (PP 30RCF 3MAPP).
Accordingly, the anisotropy ratio in the latter specimens is larger with an average value $\bar{\chi}=1.52$, whereas the samples without coupling agent exhibit $\bar{\chi}=1.46$.
As standard deviations, $\sigma_{\chi}=0.125$ and $\sigma_{\chi}=0.077$, respectively, have been determined.

The eight cracks in the PP 30RCF 3MAPP specimens deflect uniformly into the half-plane above the notch, showing a distinct variance. 
The cracks in the PP 30RCF samples behave differently.
While four cracks basically follow the paths of the PP 30RCF 3MAPP specimens, however with slightly less deflection and less scatter, two crack paths depart distinctly from the others.
In the following paragraphs, these issues will be investigated closely by simulations accounting for stochastic aspects of the fiber--matrix compound.

\subsection{Simulation of crack paths} \label{subsec:sim}

The loading analysis of the crack tip is performed by means of the $J$-integral according to Eq.~(\ref{eq:5}) based on data provided by the FEM.
The remote integration contour $\Gamma_0$, see Fig.~\ref{Fig11}, runs at some distance along the boundary of the specimen.
Crack growth is numerically simulated in conjunction with an adaptive remeshing algorithm \cite{judt3,judt4,judt2}, accounting for local field gradients and specific requirements of the integration of Eq.~(\ref{eq:6}) along the crack faces $\Gamma^{+/-}$.
Crack growth increments $\Delta a$ are chosen sufficiently small to guarantee convergence of the predicted crack paths.
In the simulations of this section, $\Delta a=0.5$mm turned out to be appropriate with regard to accuracy and computational cost.
The deflection angle $\varphi$ of an increment is determined from Eq.~(\ref{eq:14}), further checking the inequality of Eq.~(\ref{eq:12}) to identify the maximum of $J_Q(\varphi)$.
The orientation of fibers $\gamma=\gamma_{\text{loc}}-\alpha$, see Fig.~\ref{Fig0}, is either chosen deterministically and typically constant in the whole specimen, i.e. $\gamma=\mu_\gamma$, or sampled randomly according to the model in Sect.~\ref{ssec:stoch} on discrete vertices of a grid, prior to predicting one crack path.
The squarish grid with a mesh size of 2mm spans the whole domain of the crack growth simulation.
A linear interpolation yields the orientation angle $\gamma_{\text{loc}}$ of Eq.~(\ref{eq:14}) at the current position of the crack tip.

\begin{figure}[!ht]
\centering
\subfigure[Standard deviation $\sigma_{\gamma}=10^{\circ}$]{\includegraphics[width=.47\textwidth]{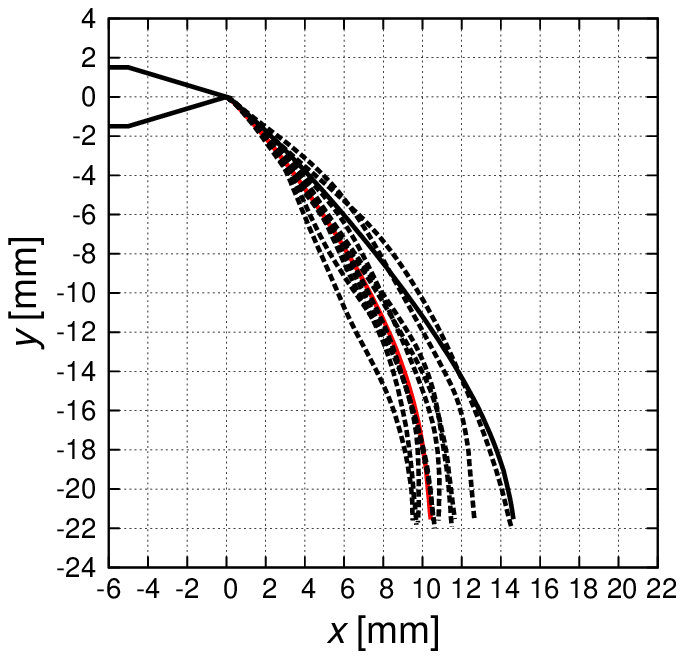}}\hfill
\subfigure[Standard deviation $\sigma_{\gamma}=20^{\circ}$]{\includegraphics[width=.47\textwidth]{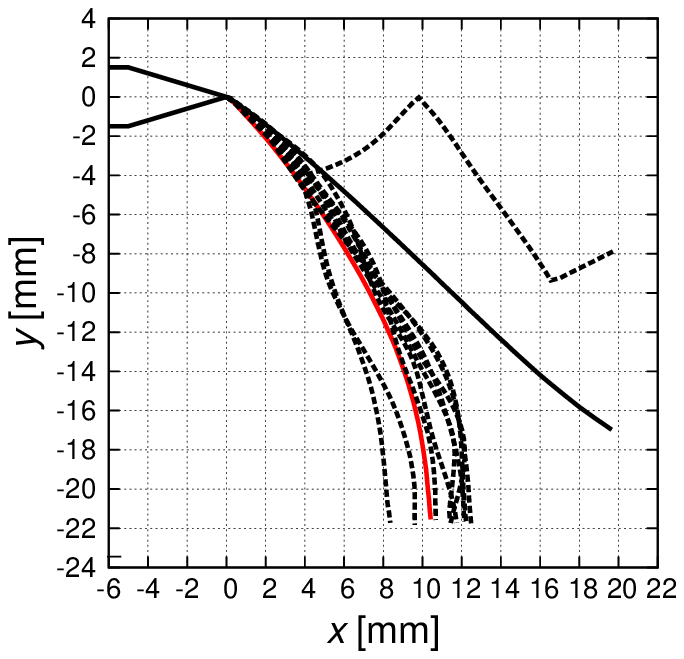}}
\subfigure[Standard deviation $\sigma_{\gamma}=30^{\circ}$]{\includegraphics[width=.47\textwidth]{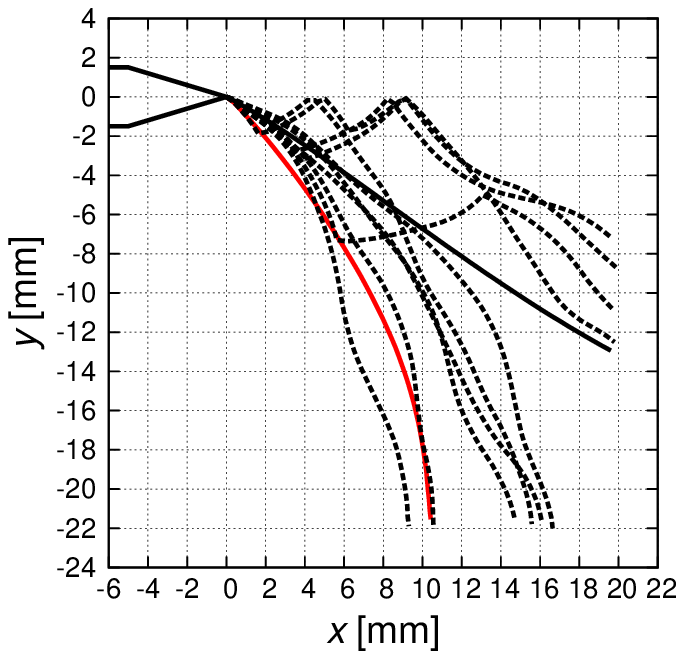}}
\caption{Crack paths from simulations with mean fiber orientation perpendicular to notch (dashed lines, $\mu_\gamma=90^{\circ}$) and different standard deviations $\sigma_{\gamma}$; correlation parameter $\ell=2.5$mm, anisotropy ratio $\chi=1.46$, interpolation factor $n=3.84$; the solid lines result from deterministic fiber orientations $\gamma=\mu_{\gamma}=90^{\circ}$ (red) and $\gamma=\mu_{\gamma}-\sigma_{\gamma}$ (black)}
\label{Fig2}
\end{figure}

To study the influence of the standard deviation of the PD of fiber alignment, ten fiber configurations of a specimen have been taken as a basis to simulate crack paths initiating from a V-shaped notch, see Fig.~\ref{Fig2}.
The expected angle of PD being $\mu_\gamma=90^{\circ}$ throughout the domain of simulation, the notch is thus perpendicular to the principal fiber orientation.
The anisotropy ratio $\chi=1.46$, taken from experiments with CT-specimens of RCF reinforced PP \cite{judt2,judt1}, is above the point of bifurcation and the correlation parameter in Eq.~\eqref{eq:SE} has been chosen $\ell=2.5$mm based on average fiber lengths of approximately 1mm, see Sect.~\ref{sec:CT}. 
Different interpolations of the anisotropic crack growth resistance $J_C (\varphi^{\ast})$, see Fig.~\ref{Fig0}, have been investigated in \cite{judt1}, whereupon an interpolation factor $n=3.84$ could be identified as most suitable. 
Initially being subject to a pure mode-I loading, the deflection gives rise to a mixed-mode state during crack growth.
The standard deviations in Fig.~\ref{Fig2} (a), (b) and (c) are $\sigma_{\gamma}=10^{\circ}$, $20^{\circ}$ and $30^{\circ}$, respectively. The solid red and black lines represent the paths of deterministic simulations with constant fiber orientations.
While the former with $\gamma=\mu_\gamma=90^{\circ}$ are identical in all three plots, the latter are based on $\gamma=\mu_\gamma-\sigma_{\gamma}$.

The stochastic crack paths for the smallest magnitude of standard deviation basically follow the deterministic paths, spanning a prediction region being slightly larger than the range indicated by the solid lines.
For larger standard deviations, individual cracks exhibit almost perpendicular deflections in the course of their growth, occurring twice or even three times.
In these cases there is a distinct departure from the deterministic paths.
Conditions allowing for crack kinking depend on the mixed-mode ratio and the local orientations of fibers with respect to the crack faces, which will be outlined in detail below.

\begin{figure}[!ht]
\centering
\subfigure[Correlation parameter $\ell=0.1$mm]
{\includegraphics[width=.47\textwidth]{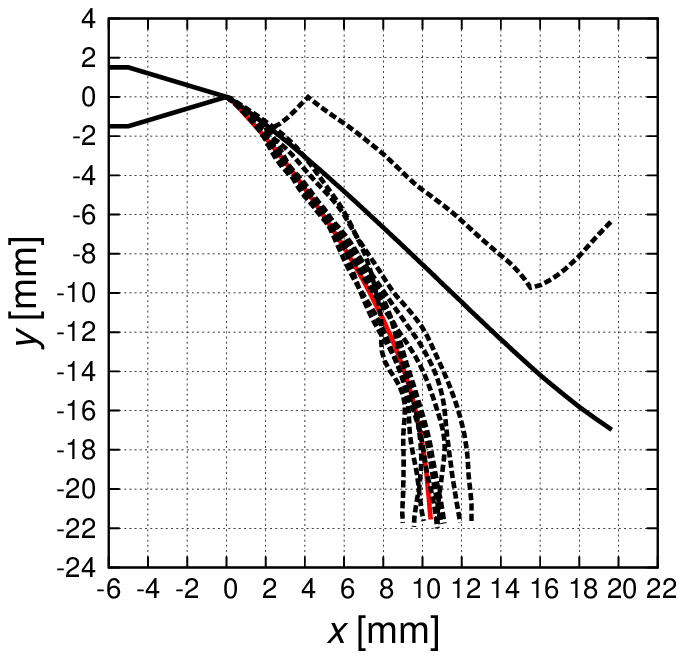}}\hfill
\subfigure[Correlation parameter $\ell=5$mm]
{\includegraphics[width=.47\textwidth]{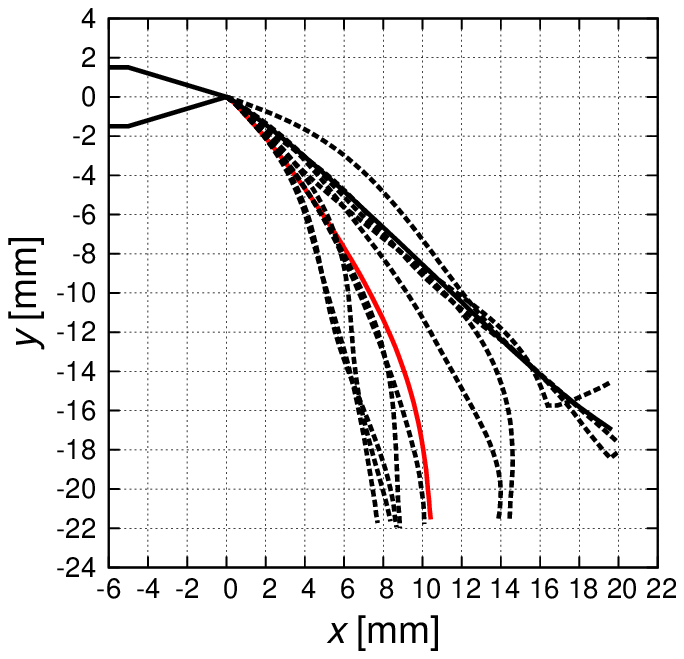}}
\caption{Crack paths from simulations with mean fiber orientation perpendicular to notch (dashed lines, $\mu_{\gamma}=90^{\circ}$) and different correlation parameters $\ell$; standard deviation $\sigma_{\gamma}=20^{\circ}$, anisotropy ratio $\chi=1.46$, interpolation factor $n=3.84$; the solid lines result from deterministic fiber orientations $\gamma=\mu_{\gamma}=90^{\circ}$ (red) and $\gamma=\mu_{\gamma}-\sigma_{\gamma}=70^{\circ}$ (black)}
\label{Fig3}
\end{figure}

In Fig.~\ref{Fig3} two different correlation parameters $\ell=0.1$mm and 5mm are investigated with a standard deviation $\sigma_{\gamma}=20^{\circ}$, which was motivated by analyses in the $\mu$--CT, see Sect.~\ref{sec:CT}, and otherwise the same parameters as in Fig.~\ref{Fig2}.
The solid red and black lines again depict deterministic simulations with $\gamma=\mu_\gamma=90^{\circ}$ and $\gamma=\mu_\gamma-\sigma_{\gamma}=70^{\circ}$.
The lower value $\ell=0.1$mm represents the limiting case of vanishing correlation, thus the fibers randomly scatter in the vicinity of the crack tip.
Accordingly, the crack paths in Fig.~\ref{Fig3} (a) exhibit comparatively small deviation from the deterministic path of $\gamma=\mu_\gamma=90^{\circ}$.
In Fig.~\ref{Fig3} (b), on the other hand, the large correlation parameter reduces the local scattering of fiber orientations, finally allowing for a larger variance of crack paths, a few of them approaching the black line from deterministic prediction for $\gamma=\mu_\gamma-\sigma_{\gamma}=70^{\circ}$.
Furthermore, the sharp deflections addressed in Fig.~\ref{Fig2} are suppressed with larger correlation parameters.

\begin{figure}[!ht]
\centering
\subfigure[Mean fiber orientation $\mu_{\gamma}=0^{\circ}$]
{\includegraphics[width=.47\textwidth]{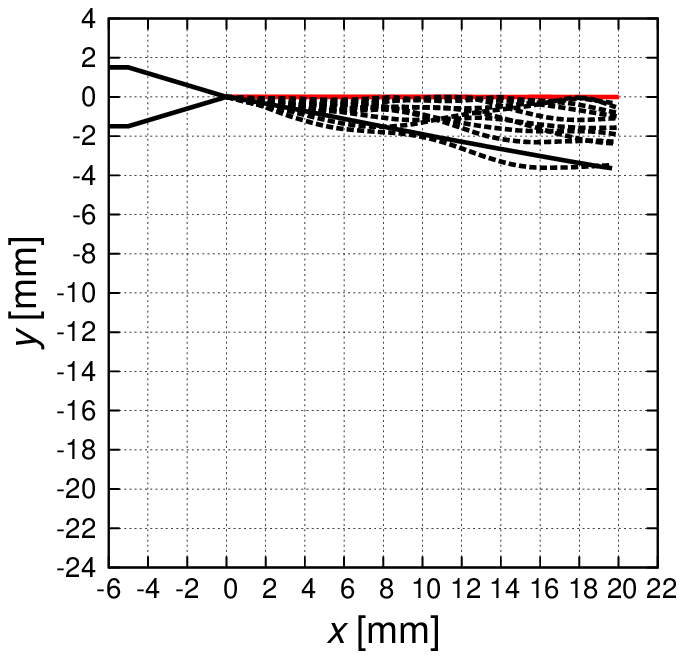}}\hfill
\subfigure[Mean fiber orientation $\mu_{\gamma}=45^{\circ}$]
{\includegraphics[width=.47\textwidth]{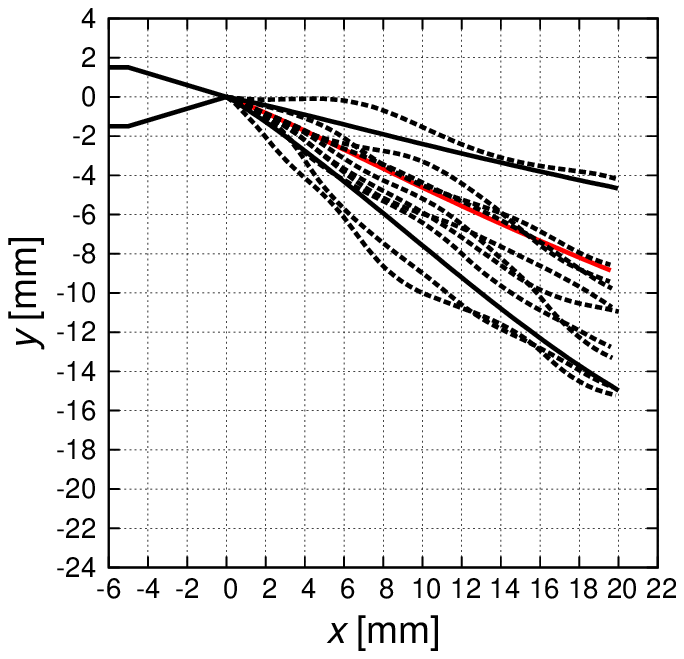}}
\caption{Crack paths from simulations with different mean fiber orientations $\mu_{\gamma}$ (dashed lines); standard deviation $\sigma_{\gamma}=20^{\circ}$, correlation parameter $\ell=2.5$mm, anisotropy ratio $\chi=1.46$, interpolation factor $n=3.84$; the solid lines result from deterministic fiber orientations $\gamma=\mu_{\gamma}$ (red) and $\gamma=\mu_{\gamma}\pm\sigma_{\gamma}$ (black)}
\label{Fig4}
\end{figure}

In Fig.~\ref{Fig4} the expected PD of fiber alignment is chosen $\mu_\gamma=0^{\circ}$ and $45^{\circ}$, respectively, throughout the domain of simulation, further taking $\sigma_{\gamma}=20^{\circ}$ and $\ell=2.5$mm.
In the former case, the crack physically deflects in the positive or negative $y$--half-plane, whereas in Fig.~\ref{Fig4} (a) cracks in the positive half-plane have been reflected to the negative one.
The solid lines once again represent deterministic simulations with constant fiber orientations $\gamma=\mu_\gamma$ (red) and $\gamma=\mu_\gamma\pm\sigma_{\gamma}$ (black).
Sharp kinks, as observed in Figs.~\ref{Fig2} and \ref{Fig3}, do not happen here, even though the scattering of crack paths is considerable in the case of $\mu_\gamma=45^{\circ}$.

\begin{figure}[!ht]
\begin{centering}
\includegraphics[width=.5\textwidth]{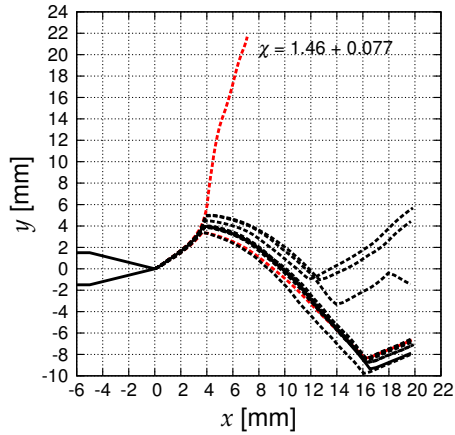}
\par\end{centering}
\protect\caption{Crack paths for one specific realization of stochastic fiber orientations ($\mu_{\gamma}=90^{\circ}$, $\sigma_{\gamma}=20^{\circ}$, $\ell=2.5$mm) and normally distributed local anisotropy ratio (black dashed lines, $\mu_{\chi}=1.46$, $\sigma_{\chi}=0.077$); red dashed lines: deterministic ratios $\chi=\mu_{\chi}\pm\sigma_{\chi}$, black solid line: $\chi=\mu_{\chi}=1.46$}
\label{Fig5}
\end{figure}

In Fig.~\ref{Fig5} the effect of stochastic fluctuations of the local anisotropy ratio is investigated for one specific realization of stochastic fiber orientations based on the parameters $\mu_{\gamma}=90^{\circ}$, $\sigma_{\gamma}=20^{\circ}$ and $\ell=2.5$mm. 
As in Fig.~\ref{FigDensity3}, a normal distribution with parameters $\mu_{\chi}=1.46$ and $\sigma_{\chi}=0.077$ extracted from experimental data \cite{judt1} is assumed.
The anisotropy ratio, according to Eq.~(\ref{eq:9}), is defined in a range $\chi>1$, which is taken into account generating stochastic values. 
In contrast to the crack path simulations of the previous figures, the parameter $\chi$ at the crack tip is randomly and in-situ determined via independent sampling in each crack growth increment.
Consequently, spatial correlations of the anisotropy ratio are not included in the calculations.
While the black dashed lines depict results of ten simulations, the red lines represent deterministic predictions in regard to the constant anisotropy ratios $\chi=\mu_{\chi}\pm\sigma_{\chi}$.
The black solid line is the crack path for $\chi=\mu_{\chi}=1.46$.

\begin{figure}[!ht]
\centering
\subfigure[fracture toughness anisotropy ratios \newline $\chi=1.46+0.077$ (red) and $\chi=1.46$ (black)]
{\includegraphics[width=.47\textwidth]{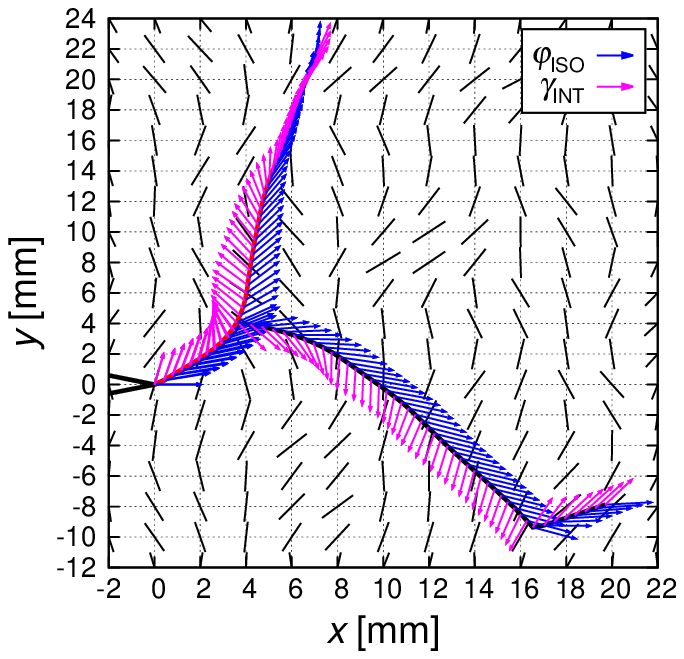}}\hfill
\subfigure[$\chi=1.46$ with stochastic (black) and deterministic (red, $\mu_{\gamma}=90^{\circ}$) fiber orientation]
{\includegraphics[width=.47\textwidth]{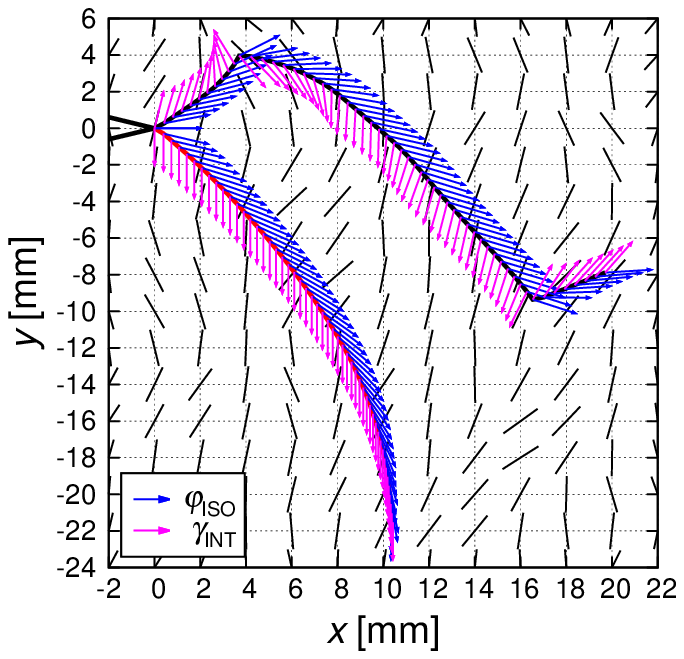}}
\caption{Stochastic fiber configuration of Fig.~\ref{Fig5} indicated by short black lines and crack paths for (a) two anisotropy ratios and (b) $\chi=1.46$ with stochastic and deterministic fiber orientation; magenta and blue arrows, respectively, represent interpolated fiber orientations and directions of the $J$-integral vector}
\label{Fig6}
\end{figure}

\begin{figure}[!ht]
\begin{centering}
\includegraphics[width=.5\textwidth]{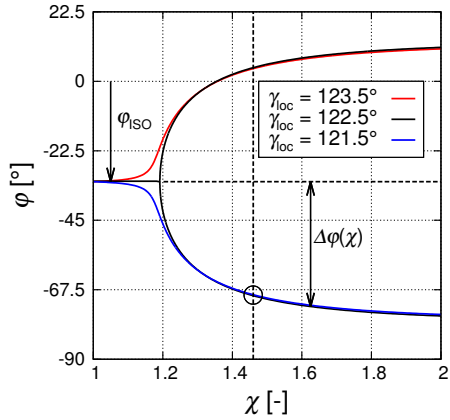}
\par\end{centering}
\protect\caption{Bifurcation diagram in the style of Fig.~\ref{Fig1} for a mixed-mode loading with three fiber orientations $\gamma_{\text{loc}}=\gamma+\alpha$ in the local crack tip coordinate system according to Fig.~\ref{Fig0}; crack angle $\alpha=65^{\circ}$, mixed-mode ratio $\Phi=0.345$ and interpolation parameter $n=3.84$ have been adopted from Fig.~\ref{Fig6} (a) at the point of crack branching}
\label{Fig7}
\end{figure}

To understand the crack paths in Fig.~\ref{Fig5}, Fig.~\ref{Fig6} (a) takes up the red dashed line for $\chi=\mu_{\chi}+\sigma_{\chi}=1.537$ and the black path for $\chi=\mu_{\chi}=1.46$.
The local PD of fiber alignment at grid points is indicated by thin solid lines distributed in the $x$-$y$--plane.
The magenta arrows illustrate the interpolated fiber orientations at the crack lines, and thus the directions of least resistance.
The blue arrows indicate the hypothetical local deflections in an isotropic medium, following the $J$-integral vector and thus the path of maximization of energy release.
According to the model of crack deflection with anisotropic fracture toughness, the crack paths approximately bisect the angles spanned by the arrows due to the large anisotropy ratio, thus seeking a compromise in between the conflicting issues of energy reduction and material resistance.
The figure further enlightens the fundamental impact of the anisotropy ratio on the crack path.
Although the parameter $\chi$ varies by only $5\%$, the paths branch and distinctively diverge after a few millimeters.
A kinking is just observed for the lower anisotropy ratio, while the larger one effectuates a smooth crack path.
It should be noted at this point that, although scarcely visible, the black and red paths diverge right from the beginning and that this slight deviation obviously gives rise to the branching.
Fig.~\ref{Fig7} will finally illuminate this issue.

In Fig.~\ref{Fig6} (b) the black line of Fig.~\ref{Fig6} (a) is adopted, whereas the red line emanates from a deterministic simulation with also $\chi=1.46$ and $\gamma=\mu_{\gamma}=90^{\circ}$.
The latter crack path thus corresponds to the red one in Fig.~\ref{Fig2}. The local fiber orientations coincide with those of Fig.~\ref{Fig6} (a), whereas the excerpts of the $x$-$y$--plane relative to the notch tip differ.
The essential impact of a stochastic versus deterministic fiber orientation is obvious.
The blue and magenta arrows, however, give evidence that in both cases the crack paths follow the concept of competing energy dissipation and crack growth resistance.
It should be noted that the red path of the uniform fiber orientation could likewise be flipped horizontally to the positive $y$--half-plane.

In Fig.~\ref{Fig7} the deflection angle determined from Eqs.~(\ref{eq:14}) and (\ref{eq:19}), respectively, is plotted versus the anisotropy ratio for three different deterministic fiber orientations $\gamma_{\text{loc}}=\gamma+\alpha$, given in the local crack tip coordinate system $(\mathbf{e}_1, \mathbf{e}_2)$ according to Fig.~\ref{Fig0}.
Their magnitudes have been chosen to illuminate the kinking behavior observed in Fig.~\ref{Fig6} (a), investigating the situation right before the branching of the solutions, where $\alpha=65^{\circ}$ is the angle of the crack at its tip with respect to the global coordinate system $(\tilde{\mathbf{e}}_1, \tilde{\mathbf{e}}_2)$.
The black lines for $\gamma_{\text{loc}}=122.5^{\circ}$, exhibiting a point of bifurcation, accordingly represents a global fiber orientation $\gamma=57.5^{\circ}$.
In contrast to Figs.~\ref{Fig1} and \ref{FigBifurcationStoch}, a mixed-mode loading with a ratio $\Phi=0.345$ prevails, corresponding to the inclination angle $\alpha$ of the crack.
Consequently, the black lines are shifted to negative deflection angles, retaining the anisotropy ratio of bifurcation and the relative angle $\Delta\varphi (\chi)=\varphi(\chi)-\varphi(\chi\to 1)$.
The offset $\varphi_{\text{ISO}}=-32.5^{\circ}$ indicates the direction of crack growth in an isotropic case $\chi\to 1$ and thus of the local $J$-integral vector, see Fig.~\ref{Fig6}.
In anisotropic materials, the deflection criterion Eq.~(\ref{eq:12}) yields a relative angle $\Delta\varphi=(\gamma_{\text{loc}}+\varphi_{\text{ISO}})/2$ for $\chi\to\infty$ and $n=4$.

The blue line in Fig.~\ref{Fig7} represents the predicted deflection for the actual local orientation of the fibers, with $\gamma=56.5^{\circ}$ being close to the condition of bifurcation, however giving rise to a unique deflection indicated by the circle at $\chi=1.46$, ultimately leading to the kinking observed for the black crack line in Fig.~\ref{Fig6} (a).
The red line in Fig.~\ref{Fig7} illustrates the case of an assumed fiber orientation slightly deviating from the bifurcation condition in the other direction.
It can be concluded that stochastic fluctuations of the anisotropy ratio only have a noticeable impact on crack paths in connection with a stochastic fiber orientation, allowing for harsh deflections and kinking. 
Basically, a bifurcation occurs if the fibers and the hypothetical direction of crack growth under isotropic conditions (magenta and blue arrows in Fig.~\ref{Fig6}) form an angle of $90^{\circ}$, or in other words if $\gamma_{\text{loc}}=\varphi_{\text{ISO}}+90^{\circ}$.
The requirement of fibers being locally perpendicular to the crack at its tip thus only holds for a pure mode-I loading.

\begin{figure}[!ht]
\begin{centering}
\includegraphics[width=.5\textwidth]{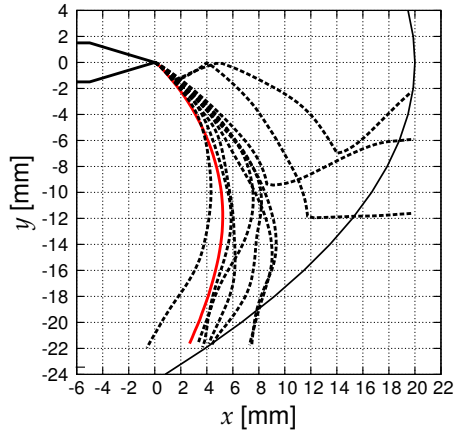}
\par\end{centering}
\protect\caption{Crack paths (dashed lines) for $\sigma_{\gamma}=20^{\circ}$, $\ell=2.5$mm, $\chi=1.46$ and $\mu_{\gamma}$ following a parabolic shape (solid black line), motivated by flow front of injection molding process; solid red line: crack path with deterministic parabolic local fiber orientation}
\label{Fig12}
\end{figure}

Finally, the parabolic shape of the flow front of the injection molding process is taken into account, assuming an expected PD of fiber alignment perpendicular to the local flow direction.
In Fig.~\ref{Fig12} the solid black line thus indicates the mean local fiber orientations $\mu_{\gamma}$ in the specimen, following the equation \cite{judt2} 
\begin{equation} \label{eq:par}
x(y)=c_0-\frac{1}{d_0} y^{2}\ ,
\end{equation}
with the parameters $d_0=30$mm and $c_0=20$mm.
The standard deviation of the stochastic simulations, with crack paths depicted by dashed lines, again has been chosen $\sigma_{\gamma}=20^{\circ}$, just as the parameters $\chi=1.46$ and $\ell=2.5$mm.
The solid red line shows the crack path from deterministic simulation with a local fiber orientation $\gamma$ following the parabolic profile.
Compared with Fig.~\ref{Fig2} (b), based on the same parameters, however, assuming a homogeneous average fiber orientation $\mu_{\gamma}=90^{\circ}$, two issues are observed.
The parabolic profile of mean orientations $\mu_{\gamma}$ is apparent in seven of the ten simulated crack paths, exhibiting a pronounced curvature.
Three paths, on the other hand, exhibit distinct departure from the deterministic crack, involving single or dual kinking, which is explained by the larger range of fiber orientation angles.

\section{Conclusions}
\label{sec:concl}

Motivated by short fiber reinforced polymer matrix composites, fracture-mechanical analyses of cracks in transversally isotropic quasi-brittle solids have been performed, with particular focus on stochastic modeling of statistical fluctuations of the transversal axis of fracture toughness.
For a pure mode-I loading and deterministic conditions with constant fiber orientations perpendicular to the crack, a threshold value of the anisotropy ratio, relevant for a bifurcation of crack deflection, is derived in closed form based on a J-vector criterion.
For mixed-mode loading conditions, a bifurcation of crack deflection prevails beyond the same threshold of anisotropy ratio, the fibers, however, in a generalized context have to be aligned perpendicular to the J-vector.
For slightly different orientations of the transversal axis a unique deflection, increasing smoothly with the anisotropy ratio, is obtained.
The stochastic model provides probability density functions and prediction regions of the deflection angle, whereupon the deterministic prediction consitutes a convex envelope for sufficiently large anisotropy ratios.
In the case of mode-I loading and perpendicular mean fiber orientations, a straight crack propagation is most unlikely, even for barely pronounced fracture toughness anisotropy.
Stochastic crack growth simulations illustrate influences of mean values and standard deviations of fiber orientations and anisotropy ratios, helping to explain the scatter of experimentally observed crack paths.

\section*{Acknowledgement}
The authors thank H.-P. Heim, J.-C. Zarges and A. Schlink for providing data of investigations in the $\mu$--CT.

%\section{Appendix}
%\label{sec:appendix}
%if required

\bibliography{article.bib}
\bibliographystyle{spbasic}

\end{document}